\documentclass[10pt,journal,comsoc]{IEEEtran}
\ifCLASSOPTIONcompsoc
  \usepackage[nocompress]{cite}
\else
  \usepackage{cite}
\fi
\ifCLASSINFOpdf
\else
\fi

\usepackage{amsmath,amssymb,amsfonts}
\usepackage{subfig}
\usepackage{array}
\usepackage{graphicx}
\usepackage{textcomp}
\usepackage{xcolor}
\usepackage[pagewise,switch]{lineno}
\usepackage{booktabs}
\usepackage{multirow}
\usepackage[ruled,linesnumbered,vlined]{algorithm2e}
\usepackage{indentfirst}
\usepackage{marvosym}
\usepackage{amsmath}
\usepackage{float}
\usepackage{soul}
\usepackage{color}
\usepackage{caption}
\usepackage{amsthm}
\usepackage{url}
\usepackage{threeparttable}

\usepackage[cmintegrals]{newtxmath}
\begin{document}

\title{Wital: A COTS WiFi Devices Based Vital Signs Monitoring System Using NLOS Sensing Model}

	\author{Xiang~Zhang,
Yu~Gu,~\IEEEmembership{Senior Member,~IEEE,}
Huan~Yan,
Yantong~Wang,
Mianxiong~Dong,~\IEEEmembership{Member,~IEEE,}
Kaoru~Ota,~\IEEEmembership{Member,~IEEE,}
Fuji~Ren,~\IEEEmembership{Senior~Member,~IEEE,}
Yusheng~Ji,~\IEEEmembership{Fellow,~IEEE}
\IEEEcompsocitemizethanks{\IEEEcompsocthanksitem Xiang~Zhang, Huan~Yan and Yantong Wang are with the School of Computer and Information, Hefei University of Technology, China. E-mail: zhangxiang@ieee.org
\IEEEcompsocthanksitem Yu~Gu (corresponding author) and Fuji~Ren are with I$^{+}$ Lab, School of Computer Science and Engineering, University of Electronic Science and Technology of China. E-mail: yugu.bruce@ieee.org, renfuji@uestc.edu.cn
\IEEEcompsocthanksitem Mianxiong~Dong
and Kaoru Ota are with the Dept. of Sciences and Informatics, Muroran Institute of Technology, Japan. E-mail: mx.dong@csse.muroran-it.ac.jp, ota@csse.muroran-it.ac.jp
\IEEEcompsocthanksitem Yusheng~Ji is with the National Institute of Informatics, Tokyo 101-8430, Japan. E-mail: kei@nii.ac.jp
}
}
\markboth{Journal of \LaTeX\ Class Files,~Vol.~14, No.~8, August~2015}%
{Shell \MakeLowercase{\textit{et al.}}: Bare Demo of IEEEtran.cls for Computer Society Journals}
\IEEEcompsoctitleabstractindextext{%
\begin{abstract}
Vital sign (breathing and heartbeat) monitoring is essential for patient care and sleep disease prevention. Most current solutions are based on wearable sensors or cameras; however, the former could affect sleep quality, while the latter often present privacy concerns. To address these shortcomings, we propose Wital, a contactless vital sign monitoring system based on low-cost and widespread commercial off-the-shelf (COTS) Wi-Fi devices. There are two challenges that need to be overcome. First, the torso deformations caused by breathing/heartbeats are weak. How can such deformations be effectively captured? Second, movements such as turning over affect the accuracy of vital sign monitoring. How can such detrimental effects be avoided? For the former, we propose a non-line-of-sight (NLOS) sensing model for modeling the relationship between the energy ratio of line-of-sight (LOS) to NLOS signals and the vital sign monitoring capability using Ricean K theory and use this model to guide the system construction to better capture the deformations caused by breathing/heartbeats. For the latter, we propose a motion segmentation method based on motion regularity detection
that accurately distinguishes respiration from other motions, and we remove periods that include movements such as turning over to eliminate detrimental effects. We have implemented and validated Wital on low-cost COTS devices. The experimental results demonstrate the effectiveness of Wital in monitoring vital signs.

\end{abstract}
\begin{IEEEkeywords}
Wi-Fi, CSI, sleep monitoring, vital signs, wireless sensing \end{IEEEkeywords}}

\maketitle

\IEEEdisplaynontitleabstractindextext

%
\IEEEpeerreviewmaketitle

\section{Introduction}
\label{Sect:int}

\IEEEPARstart{R}espiratory rate and heart rate are key physiological indicators for assessing the health of a human being, and some diseases can be detected or prevented by finding abnormalities in such vital signs, e.g., sleep apnea\cite{min2010noncontact}, asthma\cite{braun2012bridging} and sudden infant death syndrome (SIDS) \cite{facco2014sleep}. In many such diseases, patients develop symptoms only for a short period of time, and therefore, long-term continuous monitoring is needed. However, due to limited medical resources and funding, long-term inpatient observation is impractical for most people. Therefore, continuous and cost-effective monitoring of vital signs in the home setting is highly desired.


Traditional vital sign monitoring protocols are mainly based on specialized sensors attached to the body, such as electrodes for polysomnography (PSG) \cite{kushida2005practice} and electrocardiography (ECG) \cite{nappholz1992implantable}. However, these special devices are not suitable for use in the home environment since they are generally expensive and can affect sleep quality. Other methods based on attached nonspecialized sensors, such as pressure or acceleration sensors, also require contact with the body, which can cause inconvenience to users. As a result, contactless vital sign sensing has received a great deal of attention, mainly focusing on vision- and radio frequency (RF)-based solutions. However, lighting conditions limit computer-vision-based solutions. Meanwhile, although RF-based methods \cite{zhao2016emotion,yue2018extracting} can provide noninvasive vital sign monitoring without the aforementioned drawbacks, the devices (software-defined radio or radar systems) used in traditional RF-based solutions are typically costly and difficult to deploy.

Therefore, Wi-Fi-based vital sign monitoring has recently received considerable attention \cite{zhang2018fresnel, zeng2019farsense, 9076681} due to its advantages of noncontact sensing and reliance on widespread and low-cost Wi-Fi devices. The reason why Wi-Fi can be used to detect vital signs is that breathing and heartbeats causes abdominal and chest deformations, which affects the propagation of Wi-Fi signals and WiFi Channel State Information (CSI) can record this information \cite{liu2018monitoring}. In turn, we can use the recorded CSI to recover the desired vital signs.

The first challenge in Wi-Fi-based vital sign monitoring is that the torso deformation caused by respiration/heartbeats is extremely weak and only minimally affects Wi-Fi signal propagation. Therefore, a theoretical model is needed to guide the implementation of such a system. Currently, most state-of-the-art schemes are based on the Fresnel zone model \cite{wang2016human}, the Fresnel diffraction model \cite{zhang2018fresnel}, or the CSI-ratio model \cite{zeng2019farsense, zeng2020multisense}. The Fresnel zone model analyzes how human motion leads to effective displacement, and the Fresnel diffraction model \cite{zhang2018fresnel} indicates that optimal sensing performance can be achieved at or near the line of sight (LOS). However, neither of these two Fresnel-based models considers the effect of the energy ratio of LOS to non-line-of-sight (NLOS) signals on NLOS sensing. The CSI-ratio model is very useful for data processing, but it yields better results when the perception sensitivity of the deployed hardware is higher. 

The second challenge is that the effects of motions such as turning over may be mixed with those of breathing/heartbeats, thus affecting the accuracy of vital sign detection. If these movements are not distinguished from the movements of interest, they can have a detrimental effect on the monitoring results. However, current Wi-Fi-based vital sign monitoring methods all lack a method to differentiate different kinds of motions during sleep. Moreover, since breathing/heartbeats and turning over are all dynamic phenomena, existing motion segmentation schemes \cite{8047246} for use in other Wi-Fi-based sensing applications are not applicable because they aim to separate dynamic and static activities. Thus, for accurate monitoring, we need a solution that can distinguish between different types of motions.

In this paper, we propose Wital, a contactless vital sign monitoring system based on low-cost and widespread commercial off-the-shelf (COTS) Wi-Fi devices. To address the first challenge identified above, we propose an NLOS sensing model for monitoring vital signs to guide the implementation of the system. For the second challenge, we propose a regularity-based motion segmentation method that can accurately separate breathing/heartbeats from other motions. We also implement a monitoring system using MATLAB that enables easy monitoring of vital signs.

The main contributions of this paper are summarized as follows:
\begin{enumerate}
\item We propose an NLOS sensing model to investigate the relationship between the energy ratio of LOS to NLOS signals and the ability to monitor vital signs. We theoretically prove that blocking the LOS signals during NLOS sensing is beneficial for vital sign monitoring. 
\item We propose a motion segmentation method based on regularity detection, which can accurately distinguish periods of motion (such as turning over and rising from bed) that are different from vital signs.
\item We implement a vital sign monitoring system using MATLAB. Experimental results indicate that our method achieves 96.618\% and 94.708\% accuracy for breathing and heart rate detection, respectively.
\end{enumerate}

We organize the remainder of this paper as follows. In section \ref{Sect:rel}, we provide an overview of related work. We describe our NLOS sensing model in section \ref{Sect:pre}. We introduce our system design in section \ref{Sect:sys}. Then, we evaluate our method and analyze the experimental results in section \ref{Sect:per}. Finally, we conclude our work in section \ref{Sect:con}.

\section{Related Work}
\label{Sect:rel}
\begin{table*}
	\centering   	
	\caption{\label{compare} \upshape Comparison of the proposed system with the latest research results.}
	\begin{tabular}{|c|p{3cm}<{\centering}|p{5.2cm}<{\centering}|p{3.5cm}<{\centering}|p{2.3cm}<{\centering}|}
		\hline
		Reference & Vital Signs & Accuracy & Requirements &Model Support\\
		\hline
		\cite{liu2016contactless} & breathing rate (various sleep postures) & greater than 85\% & 2 transmitters and 3 receivers, 3 data streams, natural breathing& no\\
		\hline
		\cite{liu2018monitoring} & breathing rate (various sleep postures) and heart rate (only supine) & 80\% of estimation errors are less than 0.5 bpm for breathing rate, 90\% of estimation errors are less than 4 bpm for heart rate & pair of transceivers, one data stream, metronome to control breathing & no\\
		\hline
		\cite{zhang2019breathtrack} &  breathing rate (various postures) & over 99\% & pair of transceivers, cables and splitters, two data streams, metronome to control breathing & yes\\
		\hline
		\cite{zhang2018fresnel} &  breathing rate (various sleep postures) & for good positions, overall estimation accuracy is as high as 98.8\%; for bad positions, accuracy decreases to 61.5\%& pair of transceivers, one data stream, natural breathing & Fresnel diffraction model\\
		\hline
		\cite{zeng2019farsense} &  breathing rate (various sleep postures) &less than 0.3 bpm for breathing rate & pair of transceivers, two data streams, natural breathing & CSI-ratio model\\
		\hline
		\cite{liu2021wiphone} &  breathing rate (various sleep postures) & average 0.31 bpm for breathing rate & pair of transceivers, one data stream, natural breathing & ambient-reflected signal model\\
		\hline
		Our system &  breathing rate and heart rate (both for various sleep postures) & accuracy of 96.887\% for breathing rate and 94.708\% for heart rate, average errors of 0.498 bpm for detected breathing rate and 3.531 bpm for detected heart rate & pair of transceivers, one data stream, natural breathing & NLOS sensing model\\
		\hline
	\end{tabular}
\end{table*}
\subsection{Sensing with Wi-Fi}
Due to the widespread deployment of Wi-Fi devices and the convenience of wireless sensing, research on passive sensing based on Wi-Fi has received widespread attention \cite{ma2019wifi,gu2019besense,huang2020towards}. These studies are mainly based on the received signal strength index (RSSI) or CSI. The RSSI is easy to acquire, but its granularity of perception is crude. The CSI can only be obtained by modifying the underlying drivers of Wi-Fi network cards, but the granularity of sensing is better than that of the RSSI \cite{huang2021phaseanti,9075376}. 

With the help of Wi-Fi RSSI or CSI data, existing research has enabled human presence detection \cite{huang2020widet}, gesture recognition \cite{ahmed2020device,qian2017widar,8820006}, cross-domain gesture recognition \cite{zhang2021widar3,gu2022wigrunt}, localization \cite{zhao2019accurate}, sleep movement detection \cite{cao2019contactless} and driving activity detection \cite{bai2019widrive} by means of COTS Wi-Fi devices. In the past two years, Wi-Fi-based perception research has further expanded into new fields. \cite{wang2019person} used Wi-Fi devices to image key points of the human body, enabling human visualization without vision equipment. \cite{wu2020fingerdraw} used Wi-Fi devices to track symbols drawn with a finger in the air. \cite{meng2019revealing} achieved the stealing of mobile phone passwords using COTS Wi-Fi devices. By extracting the interference-independent component, PhaseAnti \cite{huang2021phaseanti} realizes the effective recognition of various activities in cochannel interference scenarios. SMARS \cite{zhang2019smars} utilizes CSI for sleep stage detection based on sleep breathing status detection and an autocorrelation function (ACF).

\subsection{Breathing and Heartbeat Monitoring}
The respiratory rate and heart rate are key physiological indicators for assessing the health of the human body. In general, the current methods used to track such vital sign information fall into three categories: contact-sensor-based, vision-based, and RF-based methods.

Most traditional solutions use contact sensors for physiological signal detection. For example, PSG \cite{kushida2005practice} and ECG \cite{nappholz1992implantable} involve measuring body functions such as breathing or heartbeats by attaching multiple sensors to a patient. H. Aly \emph{et al.} \cite{aly2016zephyr} utilized the accelerometer and gyroscope in a mobile phone to detect the breathing-induced chest motion of a person. Smart sleeping mats \cite{paalasmaa2012unobtrusive} use pressure sensor arrays for breathing detection. However, contact-sensor-based approaches are typically costly, complex to deploy, and obtrusive during measurement.

Contactless vital sign sensing methods, such as vision- and RF-based schemes are more user-friendly. However, vision-based solutions \cite{kumar2015distanceppg} are usually susceptible to variations in lighting conditions and present privacy concerns. As an alternative, research on the use of wireless radio signals as sensors has recently received increasing attention. As RF signals travel from the transmitter to the receiver, they are affected by breathing-induced chest movement along their propagation path. However, RF-signal-based solutions usually rely on special equipment, such as ultrawideband devices \cite{salmi2011propagation} or frequency-modulated continuous wave (FMCW) radar \cite{yue2018extracting, zhao2016emotion}. The devices used in these solutions are costly and not suitable for everyday environments. Compared with these solutions, Wi-Fi-based solutions are less expensive and simpler to deploy and can be implemented using COTS equipment.

Previous Wi-Fi-based breathing monitoring studies are mostly based on the RSSI \cite{patwari2013breathfinding, patwari2013monitoring, abdelnasser2015ubibreathe}. The RSSI characterizes the total received power over all paths; thus, it is a coarse-grained that is inherently incapable of capturing multipath effects. In contrast, CSI can effectively capture fine-grained channel and multipath information. Therefore, most of the latest related schemes are based on CSI \cite{ liu2016contactless, liu2018monitoring, wang2016human, zhang2018fresnel, wu2015non, chen2017tr, zeng2019farsense, zhang2019smars,wang2017tensorbeat}. Liu \emph{et al.} \cite{liu2016contactless} obtained the respiratory rate by applying the short-time Fourier transform (STFT) to the CSI amplitude. However, this solution requires the deployment of two routers and three computers. Liu \emph{et al.} \cite{liu2018monitoring} used a pair of Wi-Fi devices to monitor the respiratory rate in different sleeping postures. However, they needed to leverage LOS conditions between the devices for heartbeat detection. \cite{zhang2019breathtrack} used the CSI phase to detect the breathing rate, using cables and splitters to connect the transmitter and receiver to eliminate phase shifts.

\cite{wang2016human} proposed the Fresnel zone model to relate the depth, location, and direction of human breathing to the detectability of respiration by examining the received signal strength in the context of the Fresnel zones. \cite{zhang2018fresnel} presented the Fresnel diffraction model to relate the target person's position to the detectability of respiration within the first Fresnel zone (FFZ) and declared that the closer the person is to the LOS, the better the monitoring of respiration is in the lying posture. These models have provided excellent insight for later research. However, the Fresnel zone model and the Fresnel diffraction model do not consider the impact of changes in the LOS-to-NLOS signal ratio on vital sign monitoring. \cite{zeng2019farsense} used the CSI ratio of two receive antennas to eliminate the phase offset and utilized complex plane projection to achieve long-distance breathing detection. In this scheme, at least two antennas are required at the receiver, making it incompatible with single-antenna receivers and increasing the computational cost. Additionally, the CSI-ratio model is a data processing model. It will yield better results when the perception sensitivity of the deployed hardware is higher. The authors of \cite{liu2021wiphone} were among the first to realize respiration monitoring with cell phones, and they proposed an ambient-reflected signal model under the NLOS setting to obtain the variations in the CSI amplitude at the receiver, which varies with subtle displacements of the human chest. Their work showed that blocking the LOS signal is beneficial for NLOS sensing, but no theoretical proof was provided. We compare these systems with our work in Table \ref{compare}.

Note that this paper is an extension of our previous work that was presented at IEEE BIBM 2021 \cite{gu2021real}. In our previous work, we found that blocking the LOS signal is beneficial for NLOS sensing, and in this paper, we utilize a model based on Ricean K theory to explain why in more detail. We also propose a new motion segmentation method based on regularity detection, which can accurately distinguish sleep motions (such as turning over and rising from bed) that are different from breathing/heartbeat motions. 

\section{Preliminaries}
\label{Sect:pre}

In this section, we first analyze preliminary experiments. Then, we propose our NLOS sensing model based on Ricean K theory for better monitoring of vital signs.

\subsection{Channel State Information}
The CSI of a signal describes its attenuation along its propagation paths due to phenomena such as scattering, multipath fading or shadow fading caused by motion, and decay in power over distance. In the frequency domain, this attenuation can be characterized as follows \cite{9385792}:
\begin{equation}
	\label{equ:CH}
	Y= H \cdot  X+ N,
	\end{equation}
where $Y$ and $X$ are the received and transmitted signal vectors, respectively; $ N$ is additive white Gaussian noise; and $H$ is the channel matrix representing the CSI.

For Wi-Fi CSI, the received signal's channel frequency response (CFR) can be expressed simply as the superposition of the dynamic path CFR and static CFR:
\begin{equation}
\label{equ:CFRSUM}
H(f,t)=H_s(f,t)+H_d(f,t).
\end{equation}
The dynamic CFR can be written as:
\begin{equation}
\label{equ:CFR}
H_d(f,t)=\sum_{k\in D} h_k(f,t) e^{-j2\pi f\tau _k(t)},
\end{equation}
where $f$ and $\tau _k(t)$ represent the carrier frequency and the propagation delay on the $k^{th}$ path, respectively, and $D$ is the set of dynamic paths.
The received signal has a time-varying amplitude in the complex plane \cite{wang2016human}:
\begin{equation}
\label{equ:AMP}
|H(f,\theta)|^{2}=|H_{s}(f)|^{2}+|H_{d}(f)|^{2}+2|H_{s}(f)||H_{d}(f)|cos\theta,
\end{equation}
$\theta $ is the phase difference between the static vector and the dynamic vector, and the term that causes the amplitude fluctuation of the CSI waveform is $2|H_{s}(f)||H_{d}(f)|cos\theta $. In the case that the range and position of the motion are constant, $\theta $ is also constant, and the only factors affecting the fluctuation range are $|H_{s}(f)|$ and $|H_{d}(f)|$.

\begin{figure}
	\centering
	\includegraphics[width=0.85\columnwidth]{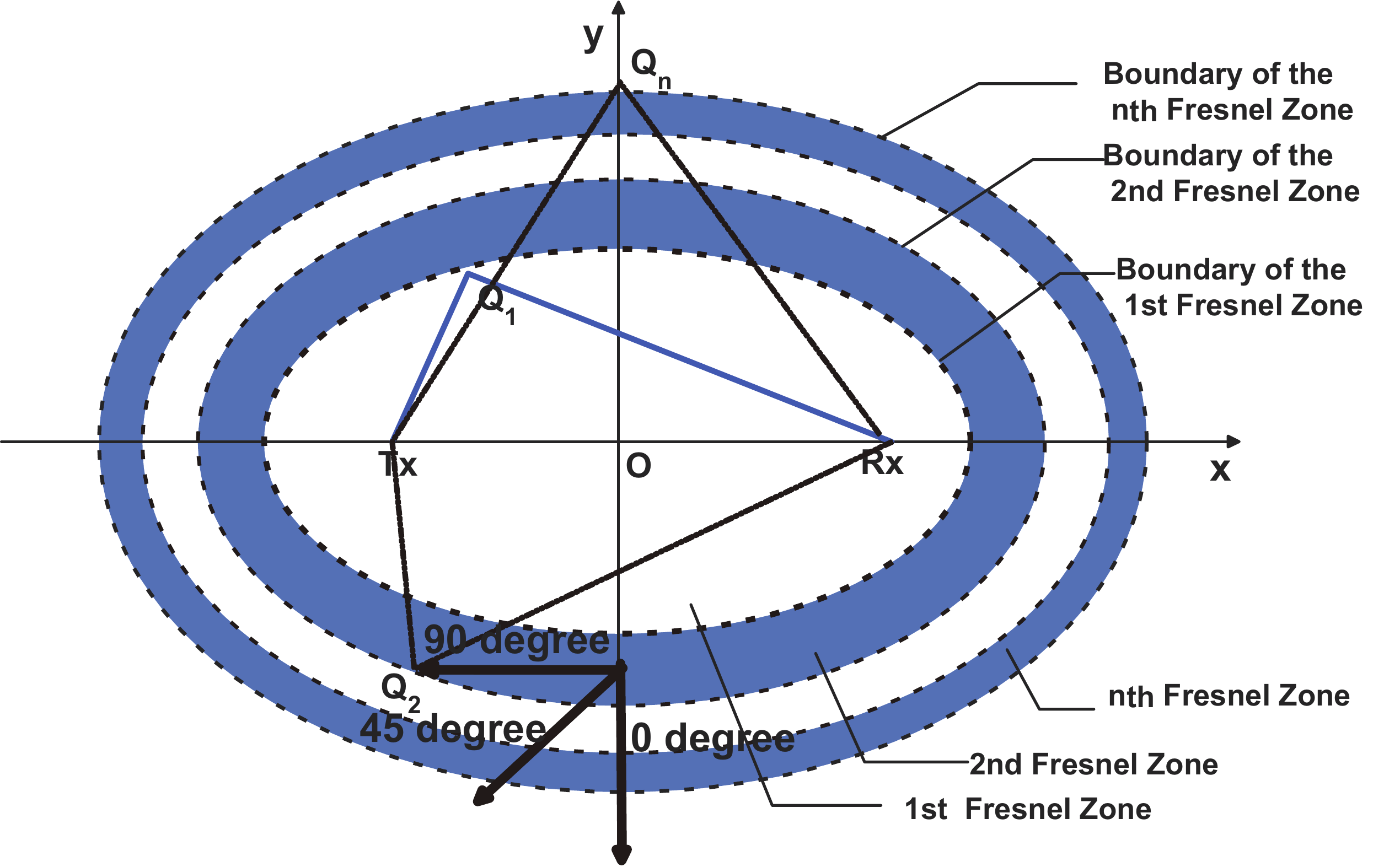}
\caption{Fresnel zones.}
	\label{fig:fz}
\end{figure}

\begin{figure*}[htb]
	\centering
\subfloat[]{\label{setting1}
\begin{minipage}{0.25\linewidth}
			\centering
			\includegraphics[width=1\textwidth]{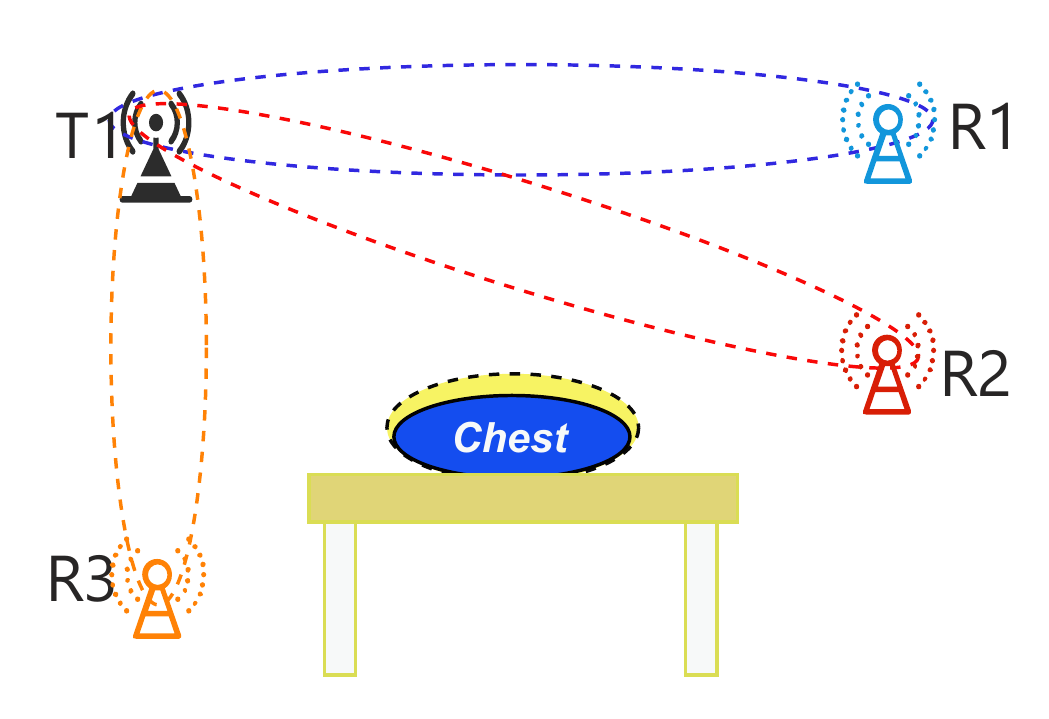}
\end{minipage}
}
\quad
\subfloat[]{\label{setting2}
\begin{minipage}{0.25\linewidth}
			\centering
			\includegraphics[width=1\textwidth]{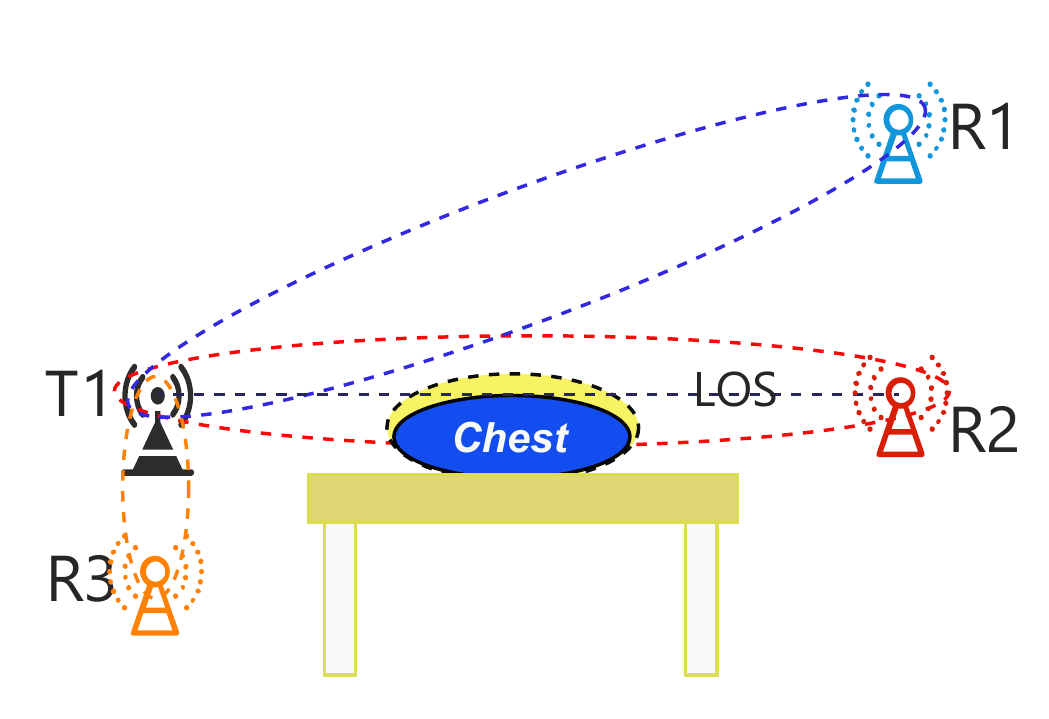}
\end{minipage}
}
\quad
\subfloat[]{\label{setting3}
\begin{minipage}{0.25\linewidth}
			\centering
			\includegraphics[width=1\textwidth]{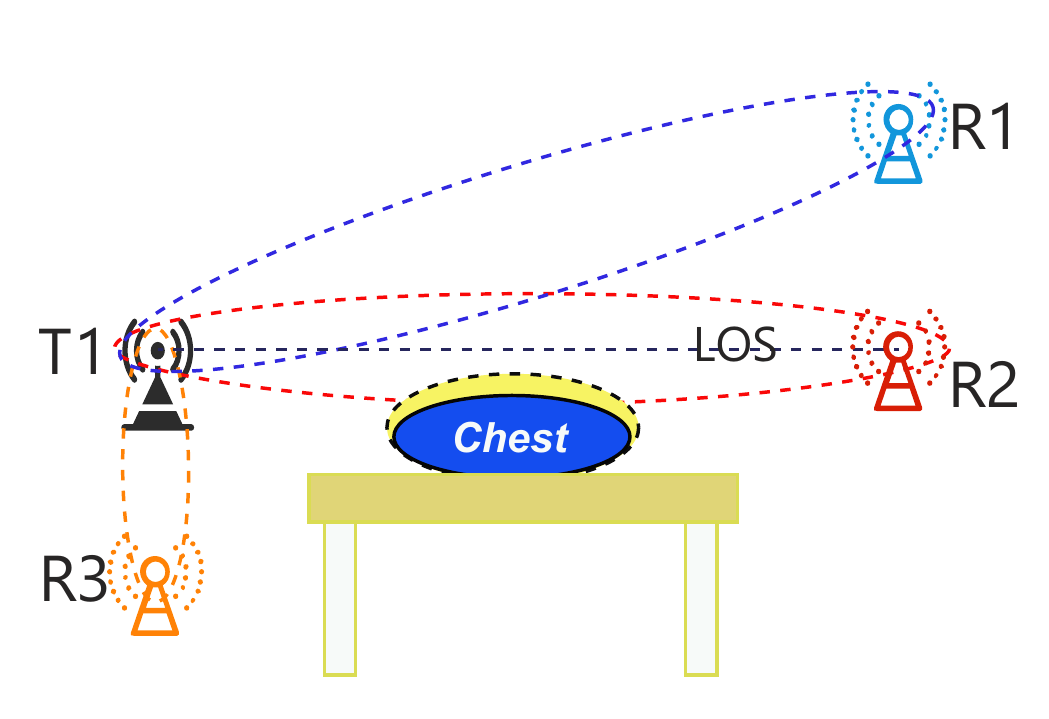}
\end{minipage}
}
\quad
\caption{Antenna settings for the preliminary experiments. The elliptical region between the antennas is the first Fresnel zone (FFZ). The region between the solid chest line and the dotted line is the range of chest deformation induced by breathing. (a) Setting 1. T1 is the transmit antenna, and R1, R2, and R3 are the receive antennas. The distance between T1 and R3 is 80 cm, and the distance between T1 and R1 is 120 cm. (b) Setting 2. The chest is on the LOS of T1--R2. The distance between T1 and R3 is 20 cm, and the distance between T1 and R2 is 120 cm. (c) Setting 3. The chest is in the FFZ of T1--R2. The distance between T1 and R3 is 30 cm, and the distance between T1 and R2 is 120 cm.}
	\label{fig:setting}
\end{figure*}

\subsection{Preliminary Experiments and Analysis}
The CSI-ratio model is focused on data processing, and it requires at least two antennas placed close together at the receiver. The ambient-reflected signal model focuses on ambient-reflected signals, which are inconsistent with our objectives. Therefore, we chose to explore the effects of the Fresnel models in our preliminary experiments. As shown in Fig. \ref{fig:fz}, the Fresnel zones are defined as a series of concentric ellipsoids, where $P_1$ and $P_2$ are the positions of the transmit antenna and the receive antennas, respectively, and $Tx$ and $Rx$ represent the transmitter and receiver, respectively. For a given radio wavelength $\lambda$, we can construct the Fresnel zones by means of the following equation \cite{zhang2019towards}:
\begin{equation} 
\label{equ:fz}
|TxQ_{n}|+|Q_{n}Rx|-|TxRx|=n\lambda/2,
\end{equation}
where $Q_n$ is a point at the boundary of the $n$th Fresnel zone.

The Fresnel zone model analyzes how human body motion causes an effective displacement $d(t)$, representing the change in the reflected (dynamic) path length of the signal. The dynamic path phase shift caused by this effective displacement can be expressed as $e^{-j2\pi f\tau _k(t)} = e ^ {-j2 \pi d (t)/\lambda} $, where $\lambda$ represents the wavelength of the Wi-Fi signal. The larger the effective displacement is, the greater the phase shift caused by the motion, and the more significant the effect on $|H_{d}(f)|$ and the CSI. Previous studies \cite{wang2016human} have shown that when the direction of human body deformation is approximately 0 degrees, the effective displacement caused by human body motion is the largest, and the sensing efficiency is the best. As the angle increases, the effective displacement decreases, and the sensing efficiency worsens \cite{wang2016human}. Moreover, when the target moves along the boundary of a Fresnel zone, there is no effect on the reflected path length.

Different from the Fresnel zone model, the Fresnel diffraction model \cite{zhang2018fresnel} focuses only on the FFZ. This mathematical model was developed to relate the location of a human target to the detectability of respiration within the FFZ by modeling various human targets as cylinders of varying size and then analyzing how a cylinder within the FFZ affects the received RF signal. The findings show that the closer to the LOS the target is, the better the respiratory monitoring effect in the lying posture. 

Based on these findings, we constructed a prototype system to carry out preliminary experiments.

\begin{figure*}[htb!]
	\centering
\subfloat[]{\label{plan1}
\begin{minipage}{0.3\linewidth}
			\centering
			\includegraphics[width=1\textwidth]{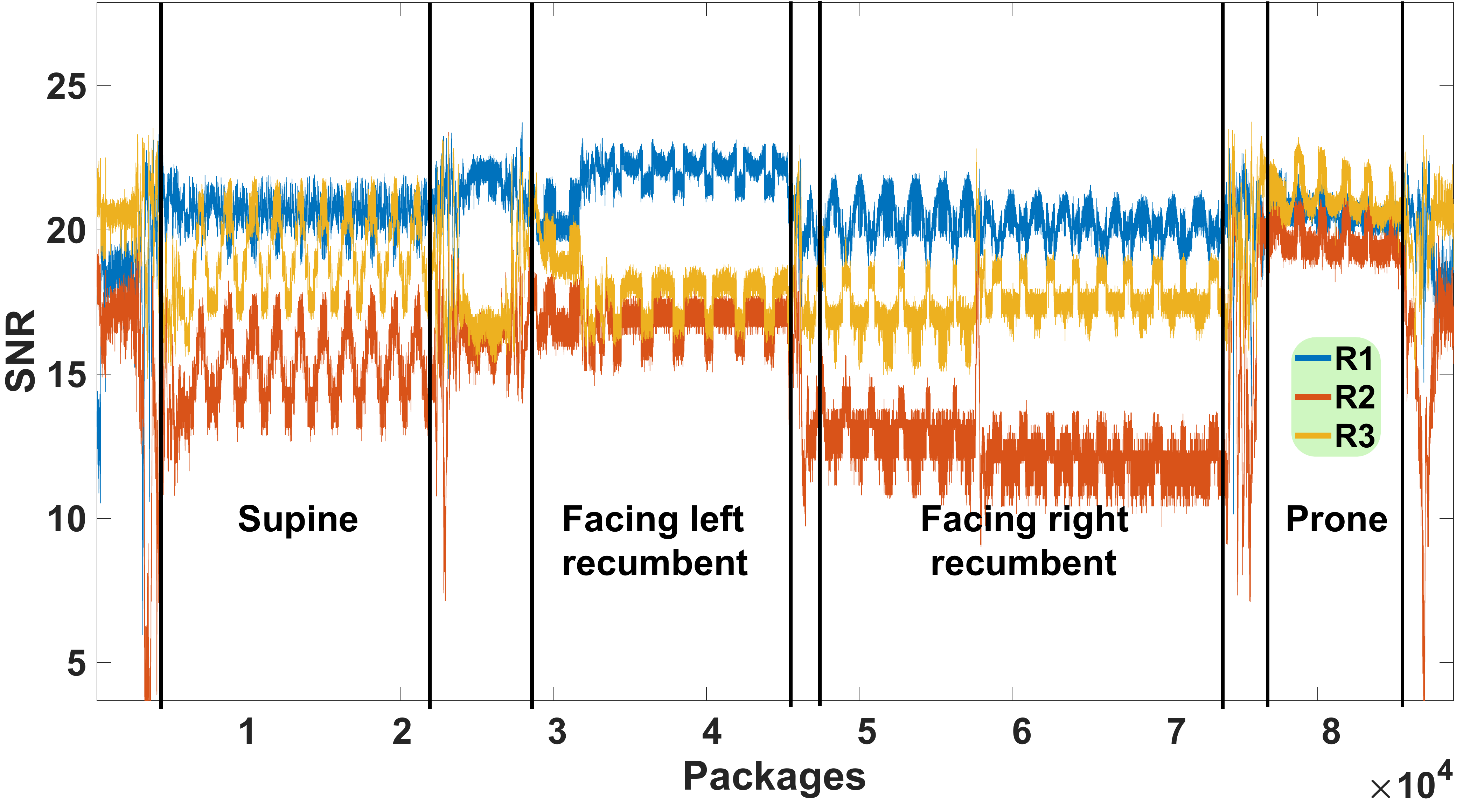}
\end{minipage}
}
\quad
\subfloat[]{\label{plan2}
\begin{minipage}{0.3\linewidth}
			\centering
			\includegraphics[width=1\textwidth]{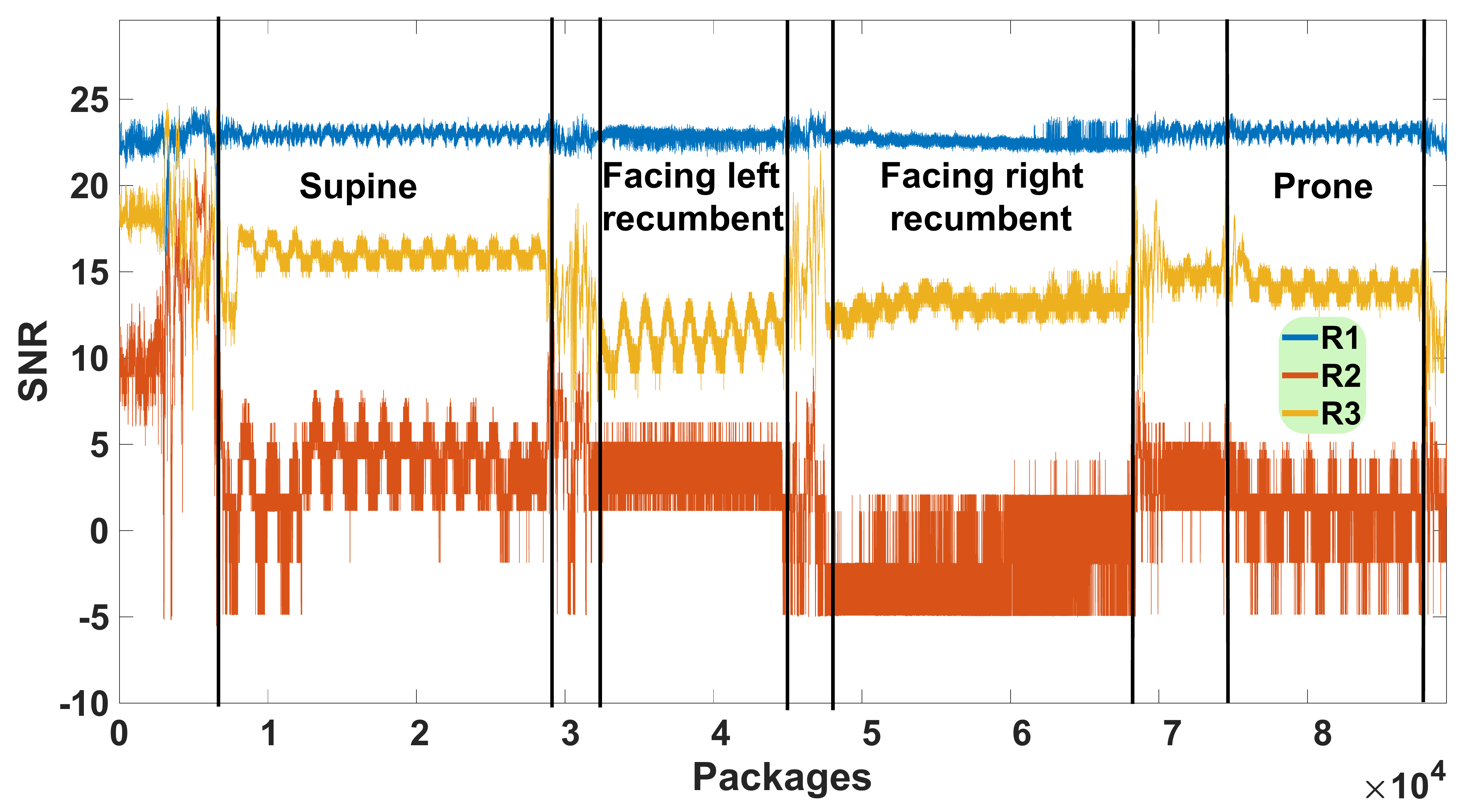}
\end{minipage}
}
\quad
\subfloat[]{\label{plan3}
\begin{minipage}{0.3\linewidth}
			\centering
			\includegraphics[width=1\textwidth]{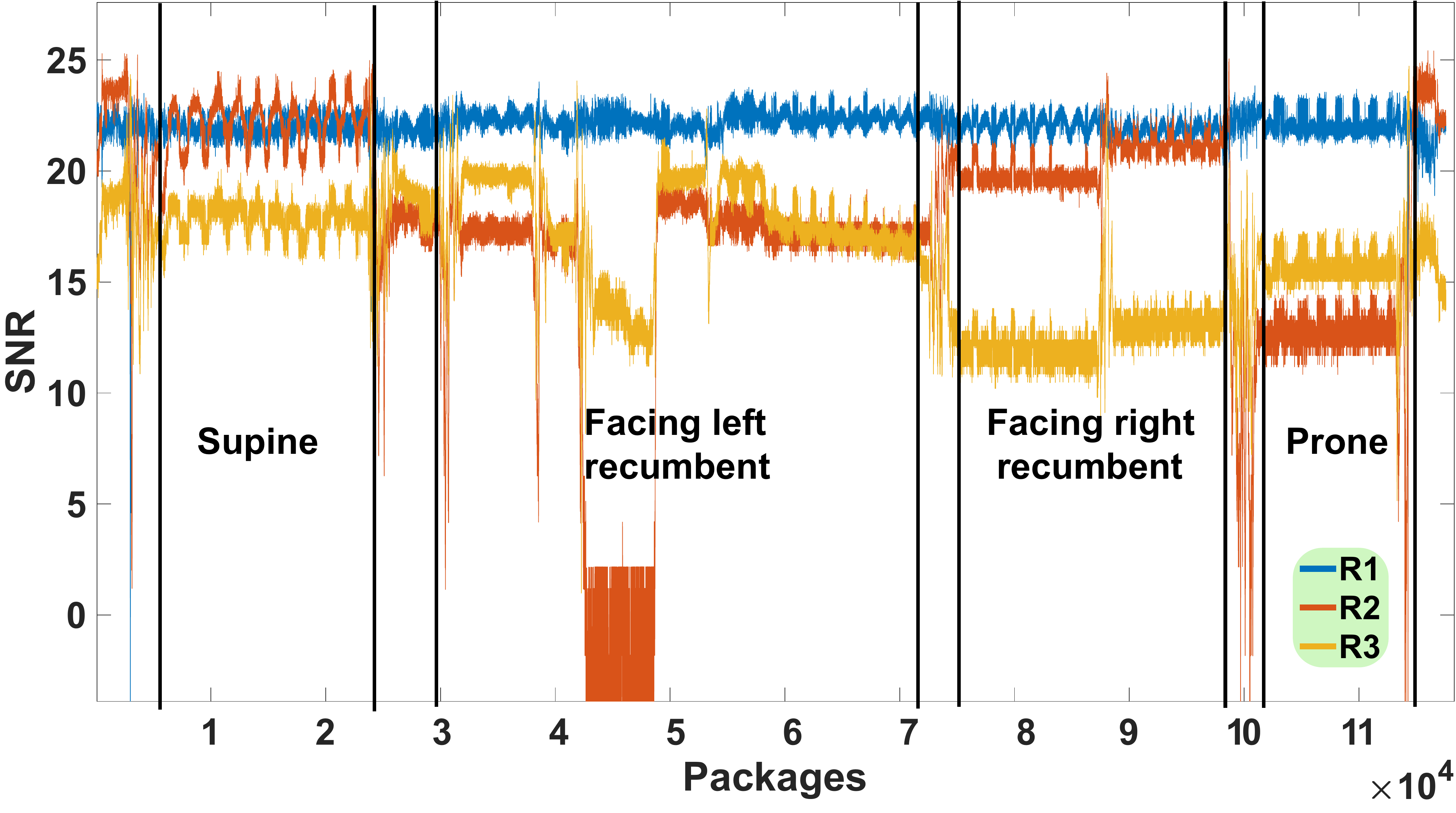}
\end{minipage}
}
\quad
\caption{Vital sign monitoring performance under different settings. (a) Performance under setting 1. (b) Performance under setting 2. (c) Performance under setting 3. SNR denotes the signal-to-noise ratio.}

	\label{fig:plan}
\end{figure*}

\noindent \textbf{[Prototype]} Our prototype system consisted of two commodity mini PCs serving as the transmitting and receiving devices. These two mini PCs were each equipped with an Intel Network Interface Controller (NIC) 5300. The antenna settings are shown in Fig. \ref{fig:setting}.

\noindent \textbf{[Participant]} One 22-year-old student participated in the preliminary experiments.

\begin{figure*}[htp]
		\centering
		\includegraphics[width=1.6\columnwidth]{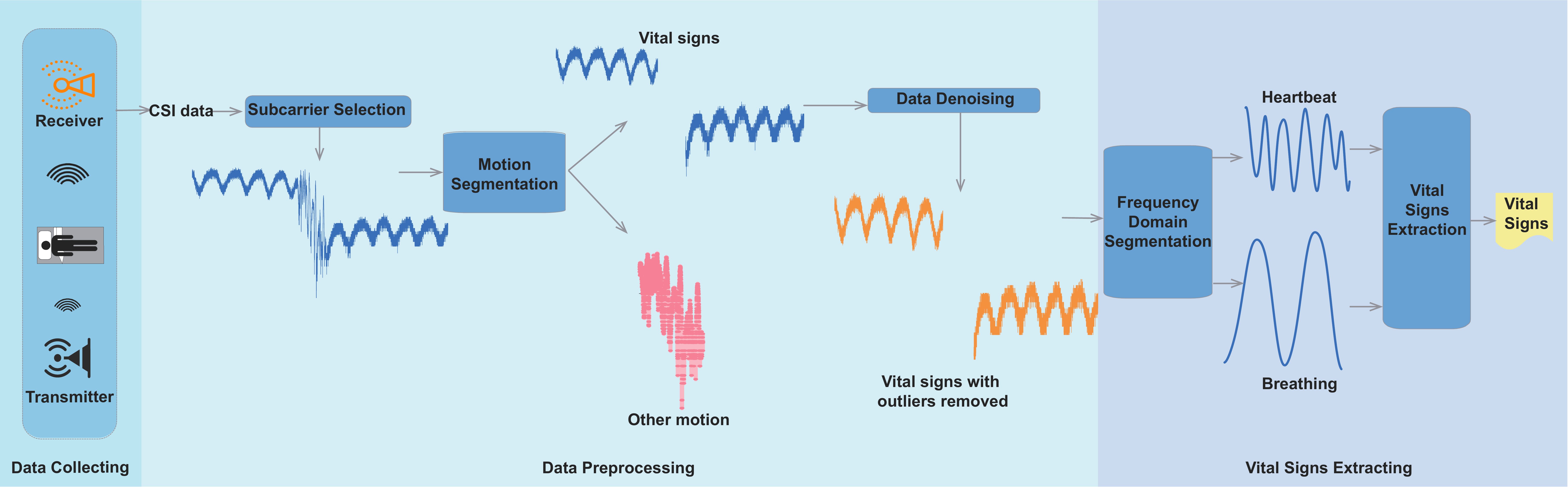}
\caption{System architecture.}
		\label{fig:system}
\end{figure*}

\noindent \textbf{[Environment]} We conducted the experiments in a $7\times 10$~m$^2$ office room with furniture, including chairs, couches, computer desks, and book cabinets, as shown in Fig. \ref{fig:system}.

\noindent \textbf{[Setting]} The package sending rate was set to 1000 Hz. The participant conducted experiments in different sleeping postures (prone, supine, left-facing recumbent, and right-facing recumbent) under different antenna settings (as shown in Fig. \ref{fig:setting}), and the experimental results are shown in Fig. \ref{fig:plan}.

By analyzing the preliminary results, we obtained the following key observations \cite{gu2021real}.

\emph{\textbf{Breathing indeed affects the channel response, and the experimental setting also affects the channel response:}}
First, we confirmed that respiration-induced signal changes were recorded under all settings. As shown in Fig. \ref{fig:plan}, we observed significant fluctuations in the CSI amplitude induced by respiration in all settings.

\emph{\textbf{Fresnel-diffraction-model-based sensing:}}
Similar to the findings presented in \cite{zhang2018fresnel}, we observed that proximity to the LOS resulted in excellent sensing performance (for both supine and prone positions in setting 2). Similar to \cite{zhang2018fresnel}, we also found that the T1--R2 antenna pair had poor perception ability when the volunteer was lying on his side. This is because the direction of trunk deformation in the anteroposterior dimension (breathing mainly causes deformation in this dimension) was roughly parallel to the LOS path of T1--R2. Accordingly, the deformation in the anteroposterior dimension could cause little significant effective displacement when the volunteer was lying on his side. The respiration-induced abdominal/thoracic deformation in the mediolateral dimension is too small; hence, its effect on signal propagation is also relatively small. Moreover, as shown in Fig.~\ref{plan2} and \ref{plan3}, it was difficult to observe significant fluctuations caused by breathing in various sleeping positions during certain periods. This is because in the sleeping state, the position of the torso differs greatly between the side-lying and flat-lying postures, and thus, it is difficult to ensure that different users have the torso close to the LOS in all sleeping postures. Therefore, it would be wise to develop a system that does not rely on the FFZ for robust monitoring of vital signs.

\emph{\textbf{Fresnel-zone-model-based sensing:}}
The Fresnel zone model \cite{wang2016human} was developed to relate the depth, location, and direction of a person's breathing to the detectability of breathing by examining the received signal strength in the context of the Fresnel zones. However, the Fresnel zone model cannot explain some of the phenomena observed in our experiments. In setting 1, when the human target is prone or supine, the direction of abdominal/thoracic deformation in the anteroposterior dimension induced by breathing is nearly parallel to T1--R3. According to the Fresnel zone model, the effective displacement caused by the thoracic/abdominal deformation is tiny, and thus, the sensing performance of T1--R3 should not be good, or at least should be worse than that of T1--R1; however, in our experiments, the sensing performance of T1--R3 was better than that of T1--R1. We believe that the reason may be that the LOS path of the T1--R3 antenna pair was blocked by the shelf. In the following subsection, we verify our conjecture through a Ricean-K-based derivation and experiments. 


\subsection{NLOS Sensing Model Based on Ricean K Theory}
To study in depth the phenomena that occurred in the preliminary experiments for better monitoring of vital signs, in this section, we propose our NLOS sensing model to analyze the relationship between the power ratio of the LOS and NLOS signals and the NLOS sensing ability based on Ricean K theory.

The Ricean K factor is defined as the ratio of the power on the LOS path to the power on the NLOS paths. The baseband in-phase/quadrature-phase (I/Q) representation of the received signal can be expressed as follows \cite{tepedelenlioglu2003ricean}:
\begin{equation}
\label{equ:RIQ}
x(t)=\sqrt{\frac{K\Omega}{K+1}} e^{j(2\Pi f_{D}cos(\theta_{0})t)+\phi_{0}}+\sqrt{\frac{\Omega}{K+1}}h(t).
\end{equation}
Here, $K$ is the Ricean factor; $\Omega$ denotes the total received power; $\theta_{0}$ and $\phi_{0}$ are the LOS angle of arrival (AOA) and phase, respectively; $f_{D}$ is the maximum Doppler frequency; and $h(t)$ is a diffuse component representing the sum of the large number of multipath components, constituting a complex Gaussian process.

Since the antennas do not move in the considered experimental scenario, i.e., $f_{D}=0$, we can simplify equation (\ref{equ:RIQ}) to:
\begin{equation}
\label{equ:RIQS}
x(t)=\sqrt{\frac{K\Omega}{K+1}} e^{\phi_{0}}+\sqrt{\frac{\Omega}{K+1}}h(t).
\end{equation}

In the case that the torso does not block the LOS signals, all LOS components and part of the NLOS components correspond to static path conditions, and the remainder of the NLOS components correspond to dynamic path conditions. Considering equation (\ref{equ:RIQS}) and ignoring the transmit power, we define $|H_{s}|$ and $|H_{d}|$ as follows:
\begin{equation}
\label{equ:HS}
|H_{s}|=\frac{K}{K+1}+\frac{1}{K+1}\cdot \rho,
\end{equation}
\begin{equation}
\label{equ:HD}
|H_{d}|=\frac{1}{K+1}\cdot (1-\rho).
\end{equation}
Here, $\rho$ is the proportion of the static path contribution to the NLOS components. Combining the above with equation (\ref{equ:AMP}), we obtain the following equation:
\begin{equation}
\begin{split}
\label{equ:HN}
|H|^{2}=|H_{s}|^{2}+|H_{d}|^{2}+2|H_{s}||H_{d}|cos\theta \\
=\frac{(K+\rho)^{2}}{(K+1)^{2}}+\frac{(1-\rho)^{2}}{(K+1)^{2}}\\ +\frac{2(K+\rho)(1-\rho)}{(K+1)^{2}}cos\theta.
\end{split}
\end{equation}

The motion-induced variation in the signal amplitude can be quantified as follows:
\begin{equation}
\label{equ:DS}
f(K,\rho)=2|H_{s}||H_{d}|cos\theta=\frac{2(K+\rho)(1-\rho)}{(K+1)^{2}}cos\theta.
\end{equation}
The output of the above formula depends on three variables, namely, $\theta$, $K$ and $\rho$. Considering that the change in the phase difference caused by breathing is relatively stable, we omit $\theta$. Then, we take the derivative of equation \ref{equ:DS} with respect to $K$ to obtain the following formula:
\begin{equation}
\label{equ:KK}
f{}'(K)=\frac{2(1-\rho )(-K^{2}-2\rho K+1-2\rho )}{(K+1)^{4}}.
\end{equation}
When $K>1-2\rho$, $f(K,\rho)$ decreases as $K$ increases. Under normal circumstances, only a small part of the signal from an omnidirectional antenna can be reflected by the human body, which means that $\rho$ is generally larger than 0.5. In other words, appropriately blocking the LOS path can make the CSI more sensitive to human body motion.

\emph{\textbf{How does $\rho$ influence the Wi-Fi sensing capability?}}

In some Wi-Fi-based sensing studies, directional antennas have been used to enhance the performance of Wi-Fi sensing. A directional antenna can transmit its signal directly toward the human body, ensuring that as much of the NLOS component as possible will be subject to dynamic path conditions (decreasing $\rho$). We can also use the proposed model to explain why this works and provide guidance for subsequent research.

We first seek a formula to explain the influence of $\rho$ on Wi-Fi sensing. We take the derivative of equation \ref{equ:DS} with respect to $\rho$ to obtain:
\begin{equation}
\label{equ:PK}
f{}'(\rho )=\frac{2(-2\rho +K-1)}{(K+1)^{2}}.
\end{equation}

When $K$ is invariant, the Wi-Fi sensing capability increases as $\rho$ increases in the interval [0,$\frac{1-K}{2}$] and decreases as $\rho$ increases in the interval [$\frac{1-K}{2}$,1]. For a more intuitive explanation, we map the CSI signal to the complex plane for further discussion.

\begin{figure}
	\centering
	\includegraphics[width=0.7\columnwidth]{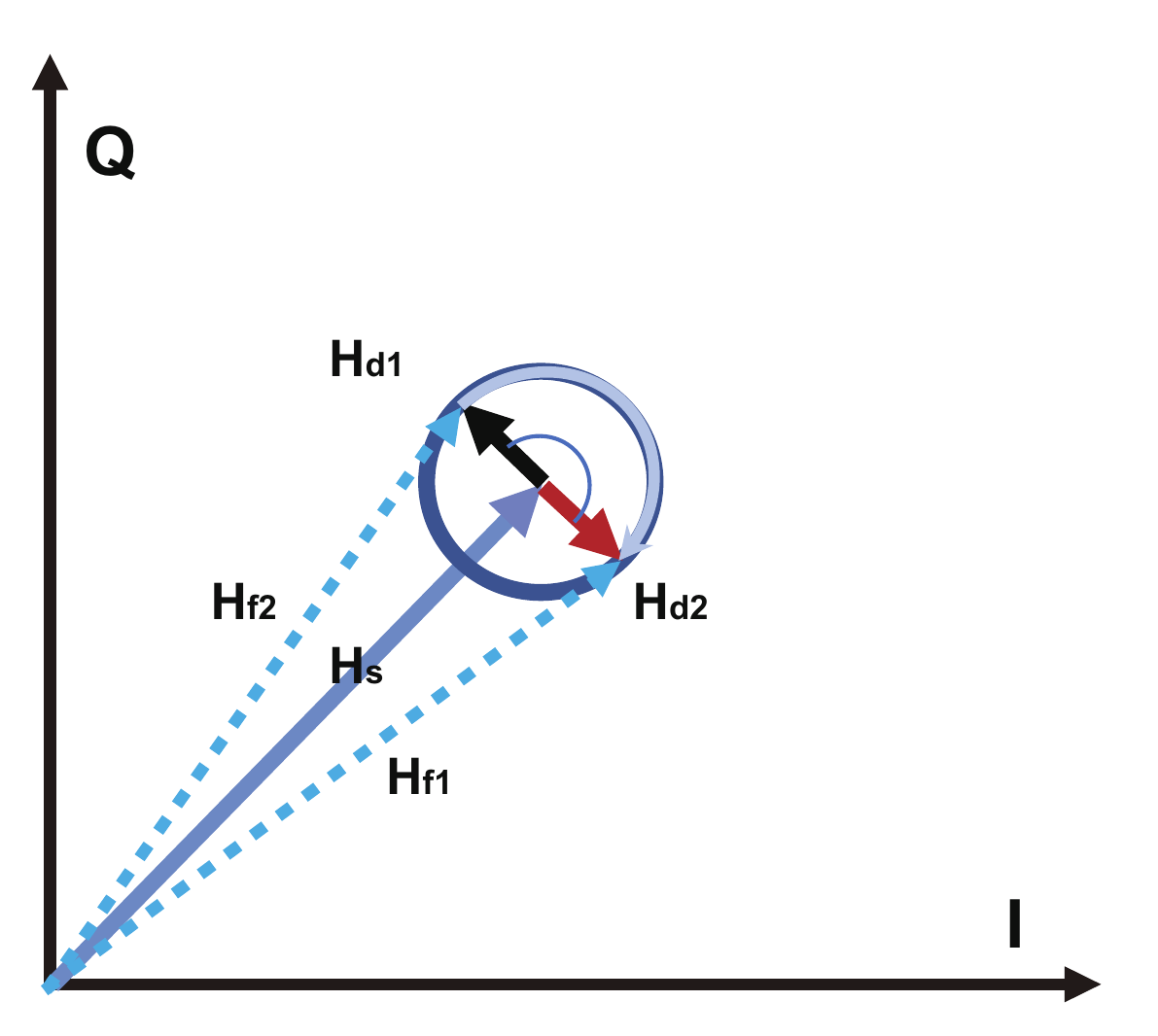}
\caption{The CSI in the complex plane when $H_{s}|>=|H_{d}|$. $H_{s}$, $H_{d}$ and $H_{f}$ represent the static and dynamic vectors and the composite vector of $H_{s}$ and $H_{d}$, respectively.}
	\label{fig:D1}
\end{figure}

Now, we can explain the conclusion in a more intuitive way. We will first explain how to use the complex plane to intuitively visualize the sensing capability \cite{zeng2019farsense}. As shown in Fig. \ref{fig:D1}, $H_{s}$, $H_{d}$ and $H_{f}$ represent the static and dynamic vectors and the composite vector of $H_{s}$ and $H_{d}$, respectively. According to \cite{zeng2019farsense}, when the dynamic path length changes on a short time scale, the amplitude of $H_{d}$ remains the same, but the phase (the angle of $H_{d}$ with respect to the $I$-axis) changes. This means that $H_{d}$ draws a circle with the end point of $H_{s}$ as its center, as shown in Fig. \ref{fig:D1}. The amplitude and phase extracted from the CSI correspond to the amplitude of $H_{f}$ and its angle with respect to the $I$-axis, respectively, and the sensing capability of the CSI can be expressed as:
\begin{equation}
\label{equ:AS}
AS=|H_{fmax}|-|H_{fmin}|,
\end{equation}
$AS$ denotes the CSI sensing ability, which is expressed as the maximum amplitude difference in the CSI waveform caused by motion. $|H_{fmax}|$ and $|H_{fmin}|$ are the maximum and minimum absolute values of the composite vector $H_{f}$, respectively.

Next, we discuss how $\rho$ affects the sensing capability in two cases: \textbf{$K> 1$} and \textbf{$0<=K<=1$}.
\textbf{When $K>1$}, $\frac{1-K}{2}<0$; according to formula \ref{equ:PK}, the sensing capability decreases monotonically as $\rho$ increases in the interval [0,1]. In the complex plane, it can be seen that $|H_{s}|>|H_{d}|$; when $|H_{s}|>=|H_{d}|$, $|H_{fmax}|^{2}=|H_{s}|^{2}+|H_{d}|^{2}+2|H_{s}||H_{d}|$ and $|H_{fmin}|^{2}=|H_{s}|^{2}+|H_{d}|^{2}-2|H_{s}||H_{d}|$. The maximum value that $AS$ can reach is $(|H_{s}+H_{d})-(|H_{s}-H_{d})=2|H_{d}|$; in other words, the larger $H_{d}$ is, the better the sensing capability.

\textbf{When $0<=K<=1$}, this means that $0.5>=\frac{1-K}{2}>=0$, and according to equation \ref{equ:PK}, the sensing capability increases as $\rho$ increases in the interval [0,$\frac{1-K}{2}$] and decreases as $\rho$ increases in the interval [$\frac{1-K}{2}$,1]. In the complex plane, when $\rho$ is in the interval [$\frac{1-K}{2}$,1], according to equations \ref{equ:HS} and \ref{equ:HD}, $|H_{s}|>|H_{d}|$, this situation is the same as in the previous paragraph: a smaller $\rho$ is better. When $\rho$ is in the interval [0, $\frac{1-K}{2}$], which means that $|H_{s}|<|H_{d}|$, the maximum value of $AS$ is $|H_{fmax}|-|H_{fmin}|=|H_{d}|+H_{s}|-(|H_{d}|-|H_{s}|)=2|H_{s}|$. In other words, the larger $H_{s}$ is, the better the sensing capability, which means that a larger $\rho$ is better.

In practical applications, it is difficult to achieve $|H_{s}|<|H_{d}|$ unless $K<1$ and most of the NLOS component is subject to the dynamic path conditions. Therefore, decreasing $\rho$ is conducive to improving the sensing capability. Considering that directional antennas can reduce $\rho$, directional antennas are concluded to be beneficial for Wi-Fi-based sensing.

\emph{\textbf{Does blocking the LOS make the sensing ability worse?}}
Blocking the LOS path reduces the Ricean K value, but does blocking the LOS make the sensing ability worse? Formula \ref{equ:DS} contains two main variables, $K$ and $\rho$. Does a situation exist in which blocking the LOS path reduces $K$ but increases $rho$, resulting in poorer motion perception? We believe that it is difficult for such a scenario to occur unless the blocking causes the static signal energy ($|H_{s}|$) that reaches the receiver to be too low. Suppose that in the worst-case scenario, the LOS path is blocked such that the NLOS signal energy increases by $L$, but no part of $L$ is allocated to the dynamic vector. Then, the amplitudes of $|H_{s}|$ and $|H_{d}|$ do not change, nor does their product, and the sensing capability is equal to that in the original case. In an actual indoor environment, it is difficult to prevent any of the blocked LOS signal from spreading to the human body at all. In summary, regardless of whether $rho$ is large, small or constant, no deterioration in the sensing capability will be caused by blocking the LOS unless the blocking severely affects the signal received at the receiver.

\section{System Design}
\label{Sect:sys}

\subsection{System Overview}
In this section, we present the system design of our vital sign monitoring system, Wital. The Wital system is shown in Fig. \ref{fig:system} and is divided into three modules:



{\bfseries \noindent Data Collection.} We collect better CSI data for vital sign monitoring based on our NLOS sensing model (blocking the LOS). Since the data collection setup is different in different scenarios, this section describes only the general data processing modules (data preprocessing and vital sign extraction) in detail. The data collection setting considered in this paper is described in the evaluation section.

{\bfseries \noindent Data Preprocessing.} We first select the best-performing subcarrier via subcarrier selection, and we then distinguish the vital signs from other motions in accordance with our motion segmentation method. Finally, we denoise the CSI data to remove outliers.

{\bfseries \noindent Vital Sign Extraction.} The preprocessed data are divided into two parts via frequency-domain segmentation: one mainly includes breathing, and the other mainly contains heartbeats. Then, we extract the breathing rate and heart rate using the Fast Fourier Transform (FFT).

\subsection{Data Preprocessing}
\label{Sect:dat}

{\bfseries \noindent Subcarrier Selection.} Different subcarriers have different central frequencies and may have different sensing performances. Therefore, it is essential to select a suitable subcarrier that can capture vital signs as effectively as possible. Based on previous experience\cite{gu2019wi}, we select the subcarrier with the largest variance for Wital.

\begin{figure}[ht]
	\centering
	\includegraphics[width=0.85\columnwidth]{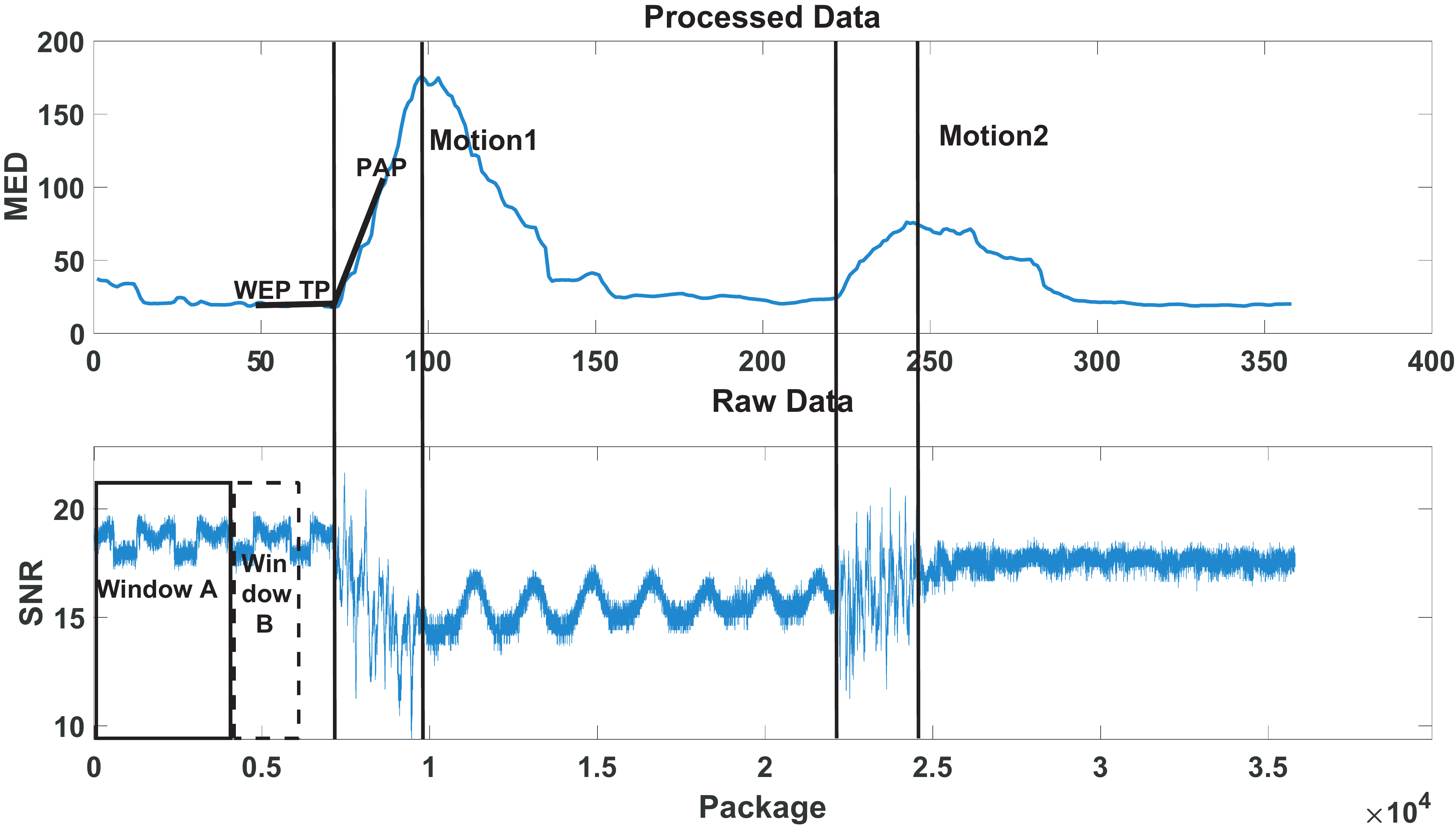}
\caption{A example of the regularity calculation results. $WEP$, $TP$, and $PAP$ represent the wave end point, turning point, and positioning activation point, respectively.}
	\label{fig:regularity}
\end{figure}

{\bfseries \noindent Motion Segmentation.}
During vital sign monitoring, it is impossible for the monitored person to remain quiescent at all times. Random movements such as turning over can affect the accuracy of vital sign monitoring; therefore, it is necessary to design a motion segmentation method to locate and filter out motions that are different from breathing/heartbeats.

Since both breathing/heartbeats and turning over are movements, the dynamic vs. static segmentation method used in previous studies is not applicable. To fill this gap, we propose a motion segmentation method that can distinguish vital signs from other movements. Our method is based on the assumption that breathing and other activities have different regularities. We choose a data sequence containing two instances of turning over as well as normal breathing as an example to verify this hypothesis, as shown in Fig. \ref{fig:regularity}.

To calculate the regularity of the CSI data, we first establish a variable-length window $A$ and a fixed-length window $B$. $A$ starts at the beginning of the CSI waveform, $B$ starts at the end of $A$, and the initial length of both $A$ and $B$ is 2000 packets. We calculate the minimum Euclidean distance (MED) between the CSI data contained in $B$ and the data in $A$ and record the result. Then, we expand window $A$ forward by 100 packets, shift window $B$ forward by 100 packets,
and calculate the MED again. The process is iterated until the end of the waveform is reached.

The results are shown in Fig. \ref{fig:regularity}. The MED increases sharply when the first instance of turning over occurs and begins to decrease once this movement ends. From the experimental results, we can observe that the regularities differ between breathing and turning over, whereas the regularities of actions of the same type are similar.

The key steps of our segmentation method are as follows:
\begin{enumerate}
\item \textbf{Initialization}. Establish a variable-length window $A$ and a fixed-length window $B$. $A$ starts at the beginning of the CSI waveform, $B$ starts at the end of $A$, and the initial length of both $A$ and $B$ is 2000 packets. Also establish an empty set $MA$.

\item \textbf{Activation point detection}. Calculate the MED of the CSI windows as described above. If $ MED > v\cdot ave (MA) $ in the current iteration (where $ave()$ is the averaging function), mark this point as a positioning activation point (\textit{PAP}) and go to the next step; otherwise, record the MED into $MA$ (where $v$ is the threshold for judging whether an action occurred; $v=2.5$ in our experiments).

\item \textbf{Start-point positioning}. Based on the obtained MED waveform ($MA$), construct an auxiliary positioning waveform to accurately locate the start point of the motion. Establish two points in front of \textit{PAP} as the turning point (\textit{TP}) and the wave end point (\textit{WEP}). The distance from \textit{TP} to \textit{WEP} is 20 packets, and the constructed waveform is as shown in Fig.~\ref{fig:regularity}. Calculate the Euclidean distance (ED) between the constructed waveform and the MED waveform and record the result; then move \textit {TP} forward by 10 packets and repeat the above process for 50 iterations. Finally, choose the \textit {TP} with the minimum ED as the start point of the motion.

\item \textbf{End-point positioning}. Set the identified start point as the beginning of the CSI waveform, initialize the parameters as described in the first step, and calculate the MED as described above. Then, use the previously described method to position the end point of the motion. Set the identified end point as the beginning of the CSI waveform, and return to step 1. Repeat the entire process until the end of the monitoring period.
\end{enumerate}

\begin{figure}
		\centering
		\includegraphics[width=0.9\columnwidth]{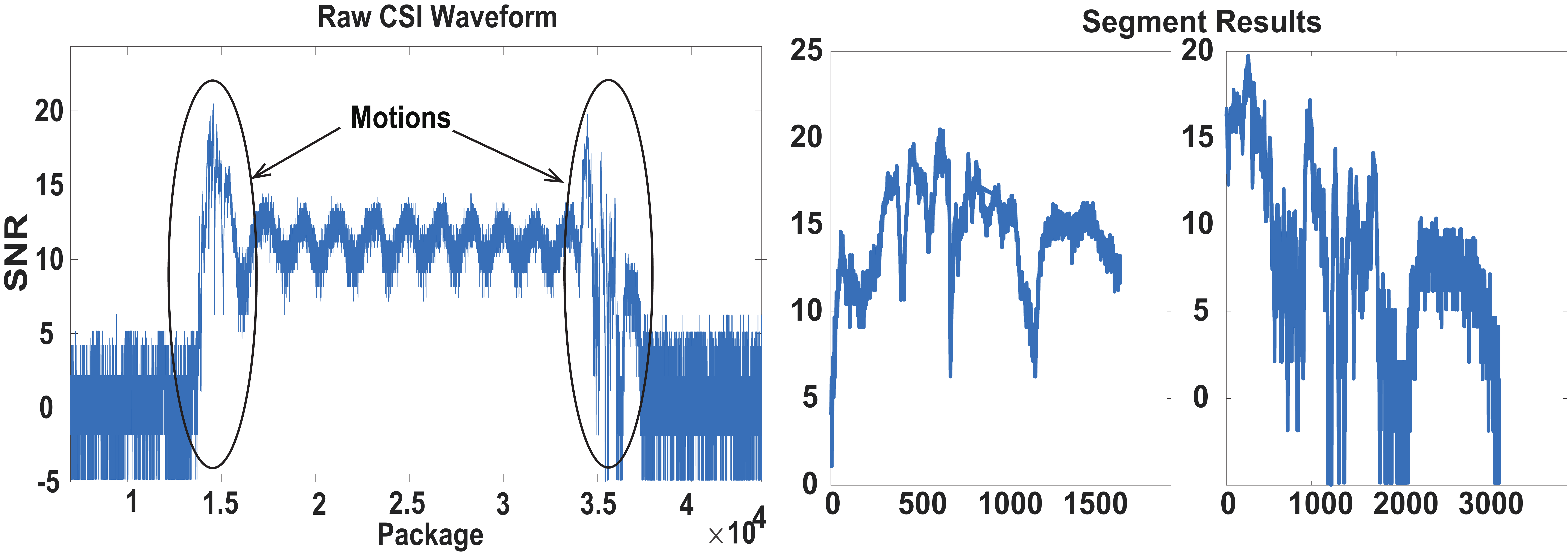}
\caption{Performance of our motion segmentation method.}
		\label{fig:segres}
\end{figure}

\begin{figure}
		\centering
		\includegraphics[width=0.85\columnwidth]{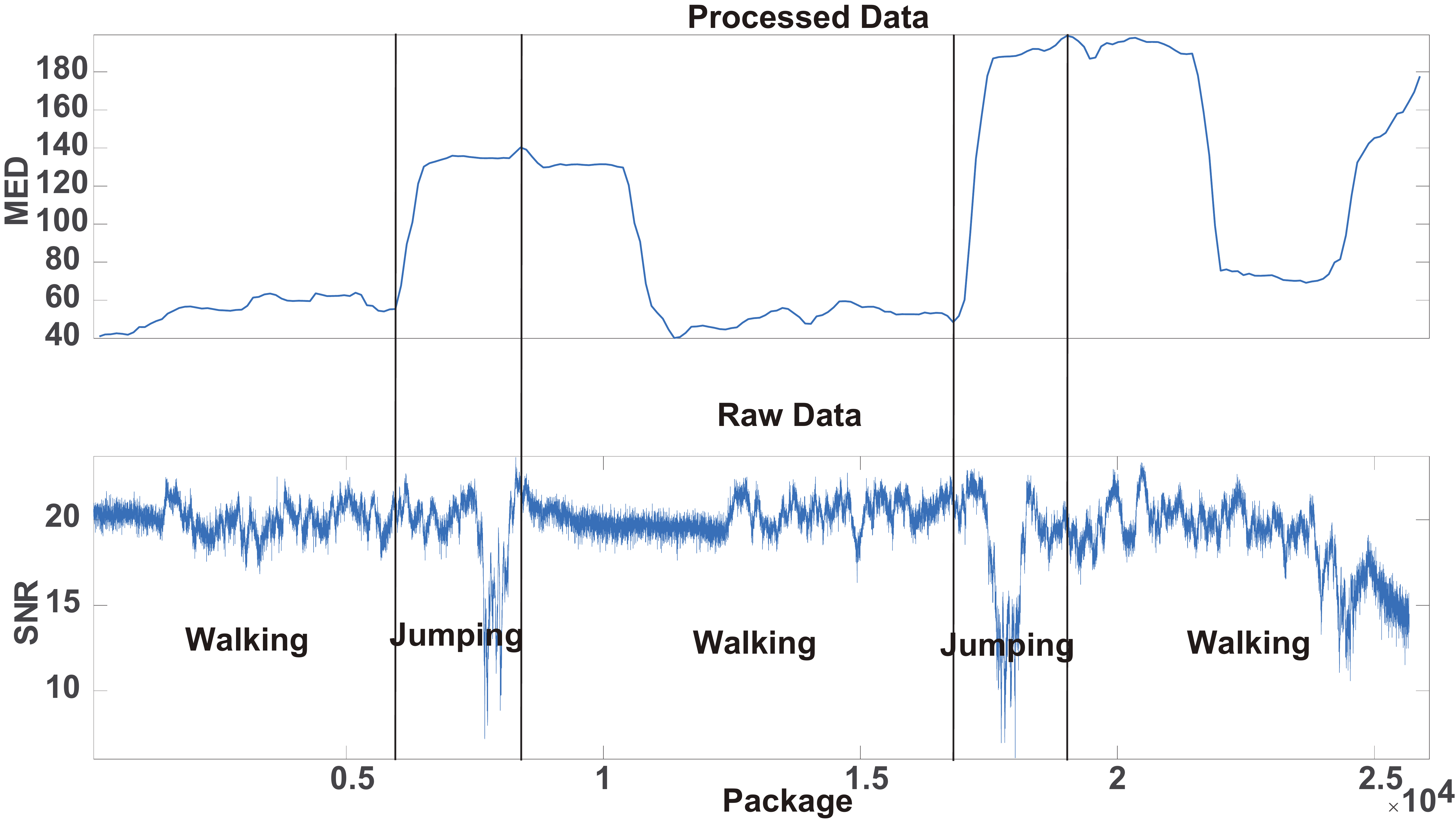}
\caption{Performance of our motion segmentation method when handling different activities.}
		\label{fig:actseg}
\end{figure}

Figs. \ref {fig:segres} and \ref{fig:actseg} show the experimental results of our proposed segmentation method. Our inspiration for designing this method is that the regularity of respiration differs greatly from those of other activities; consequently, as shown in Figs. \ref {fig:segres} and \ref{fig:regularity}, our method can precisely locate the time period during which such an activity occurs. For different pairs of activities, the difference in regularity may be smaller, and thus, the positioning of these activities may not be very precise. We tested the performance achieved in distinguishing between walking and jumping using the proposed method, as shown in Fig. \ref{fig:actseg}, from which it can be seen that the segmentation accuracy is still within an acceptable range.

{\bfseries \noindent Data Denoising.} The received CSI data may contain a large amount of interference noise due to equipment and environmental factors. In the preprocessing module, we use the Hampel filter to filter out outliers that have markedly different values from other neighboring CSI measurements. The goal of the Hampel filter is to identify and replace outliers in a given series. Specifically, we calculate the median of the set consisting of the current CSI sample and its six surrounding samples (three on each side) and use the median absolute deviation to calculate the standard deviation of the set. If the difference between the current sample and the median exceeds three times the standard deviation, this sample will be replaced by the median value.

\begin{figure}
	\centering
	\includegraphics[width=0.9\columnwidth]{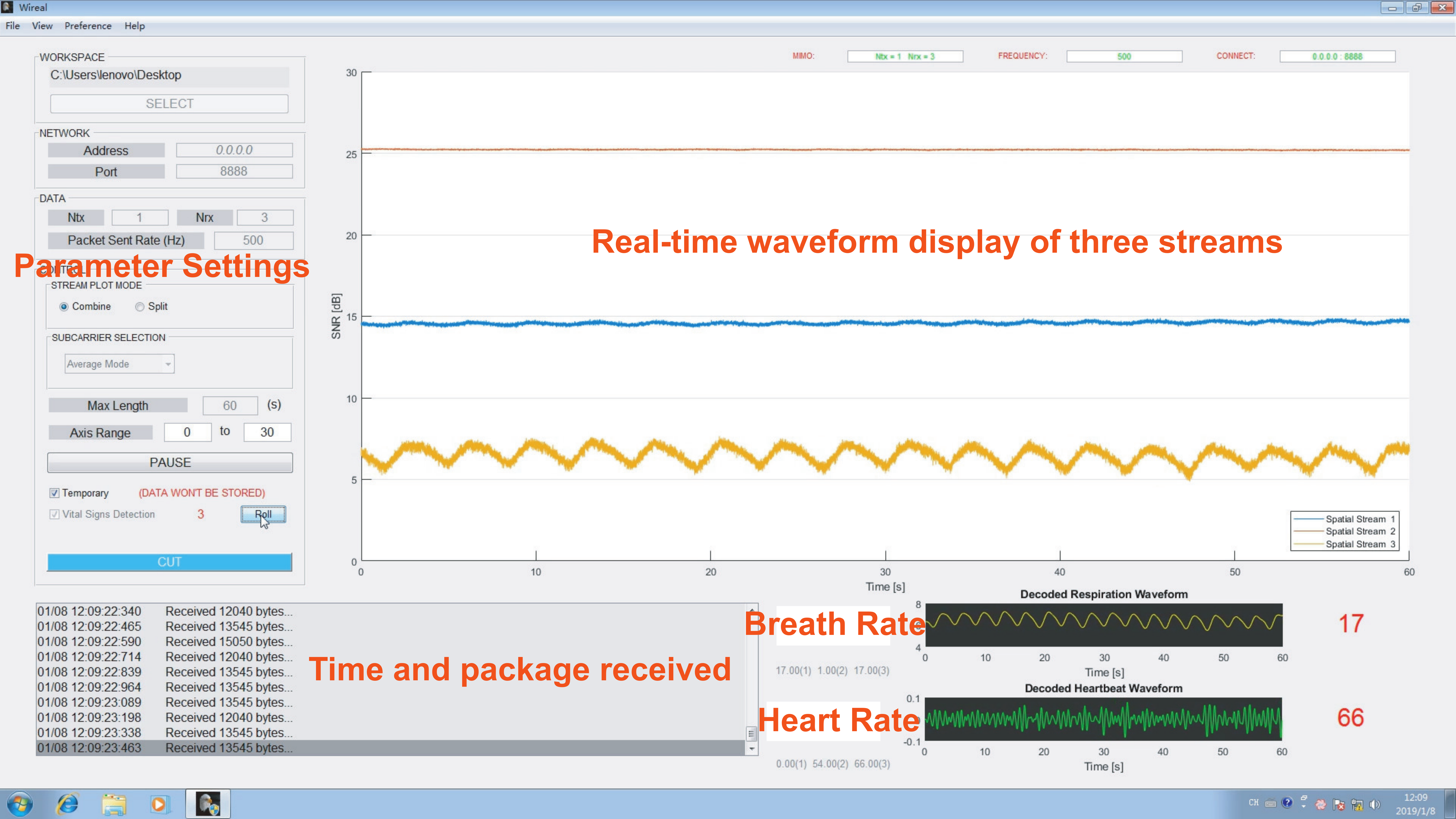}
\caption{Real-time system interface. The upper left part is the parameter setting interface, which can be used to set the storage location of the received data, the time axis scale, the receiving rate, the sender's IP address, whether to enable vital sign detection, etc. The bottom left shows the time when the current packet was received and the size of the packet. The top right is the real-time display of the CSI waveform, and the bottom right is the real-time display of vital signs.}
	\label{fig:realtime}
\end{figure}

\begin{figure*}[ht]
	\label{he}
	\centering
\subfloat[]{\label{heart}
\begin{minipage}{0.32\linewidth}
			\centering
			\includegraphics[width=1\textwidth]{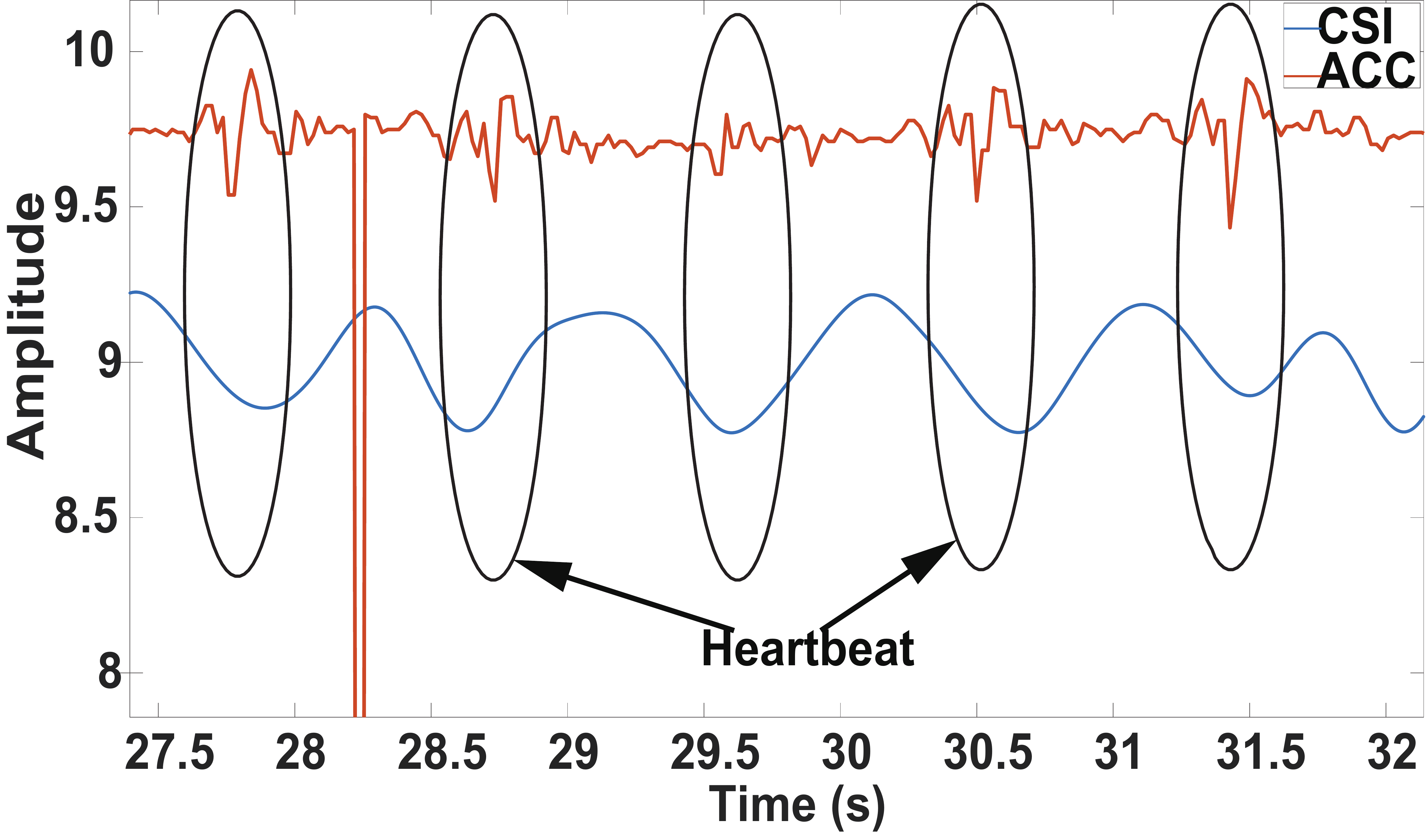}
\end{minipage}
}
\subfloat[]{\label{heartfr}
\begin{minipage}{0.32\linewidth}
			\centering
			\includegraphics[width=1\textwidth]{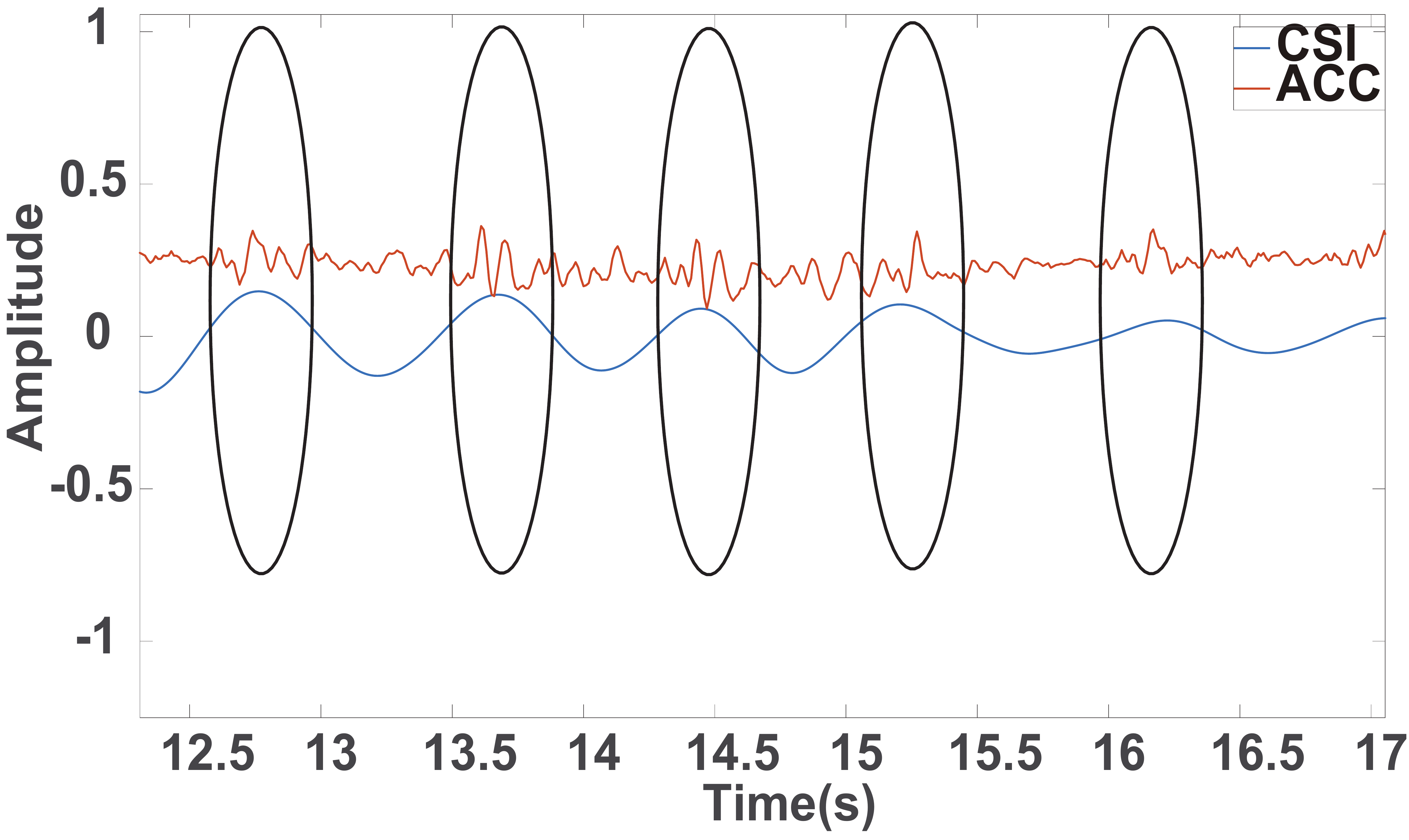}
\end{minipage}
}
\subfloat[]{\label{heartfl}
\begin{minipage}{0.32\linewidth}
			\centering
			\includegraphics[width=1\textwidth]{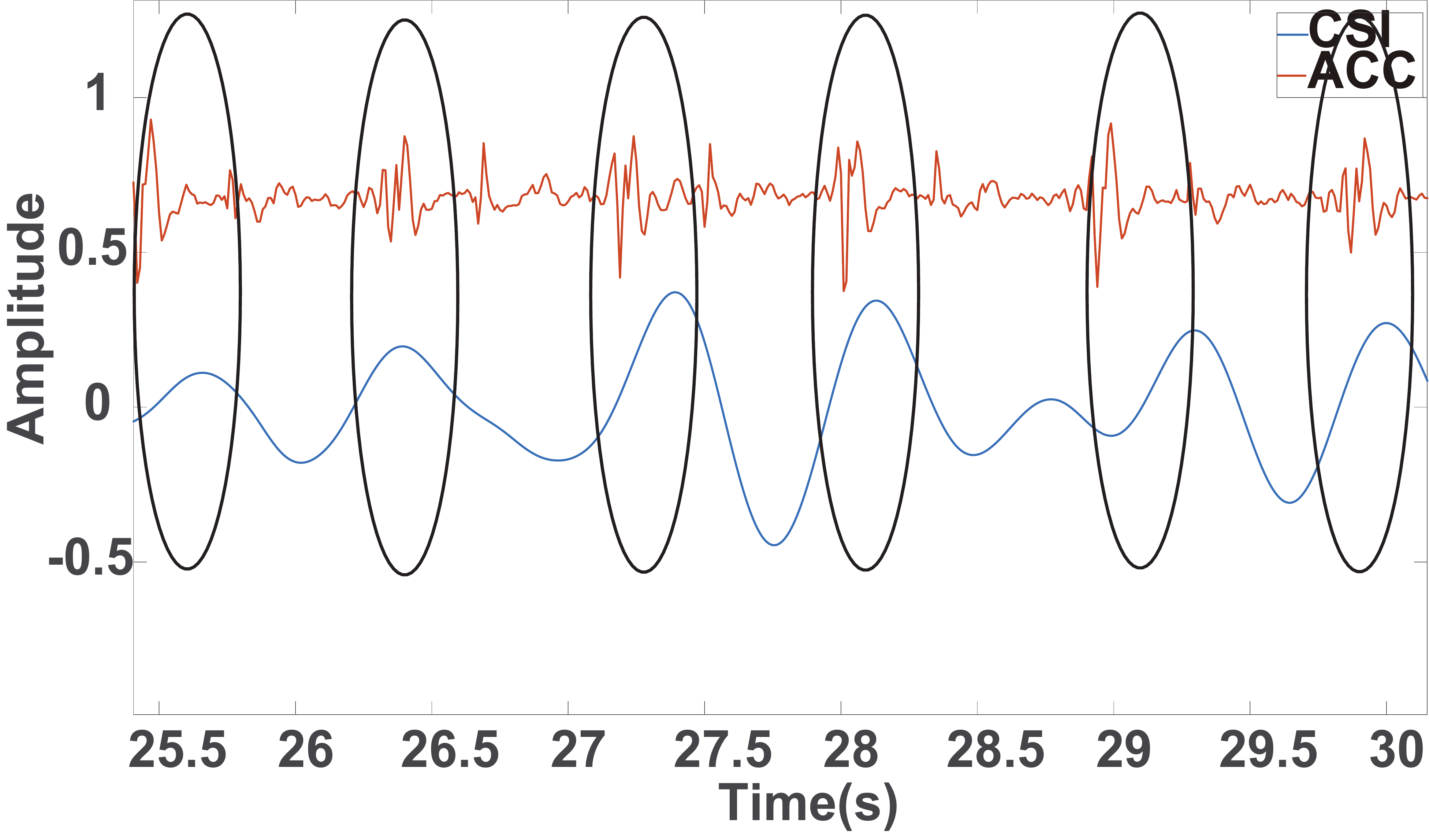}
\end{minipage}
}
\quad
\caption{Comparison of the processed CSI and the accelerometer (ACC) readings for heartbeats. (a) Supine. (b) Facing right. (c) Facing left.}
	\label{fig:heart}
\end{figure*}

\subsection{Vital Sign Extraction}

{\bfseries \noindent Frequency-Domain Segmentation.} The trunk deformation caused by the heartbeat is very small, and the changes in the CSI caused by this deformation can be overwhelmed by the changes caused by breathing \cite{liu2018monitoring}. Therefore, we first need to segment them in the frequency domain. Here, we leverage Butterworth bandpass filters based on some prior knowledge in the frequency domain: the frequency range associated with a normal heartbeat is 60 bpm to 120 bpm, corresponding to 1 Hz to 2 Hz, and the frequency range associated with normal breathing is 15 bpm to 30 bpm, corresponding to 0.25 Hz to 0.5 Hz.

{\bfseries \noindent Vital Sign Extraction.} After segmenting the CSI in the frequency domain, we extract the heart and respiratory rates by applying the FFT. We have used MATLAB to implement a real-time system for the processing and display vital signs to facilitate our experiments, as shown in Fig. \ref{fig:realtime}. This system works on the monitoring side and is connected to the data acquisition equipment via a network.

\section{Performance Evaluation}
\label{Sect:per}

In this section, we first evaluate the proposed vital sign monitoring system, Wital, and then verify the effectiveness of our NLOS sensing model.

\subsection{Experimental Setup}

\begin{figure}[ht]
	\centering
	\includegraphics[width=0.85\columnwidth]{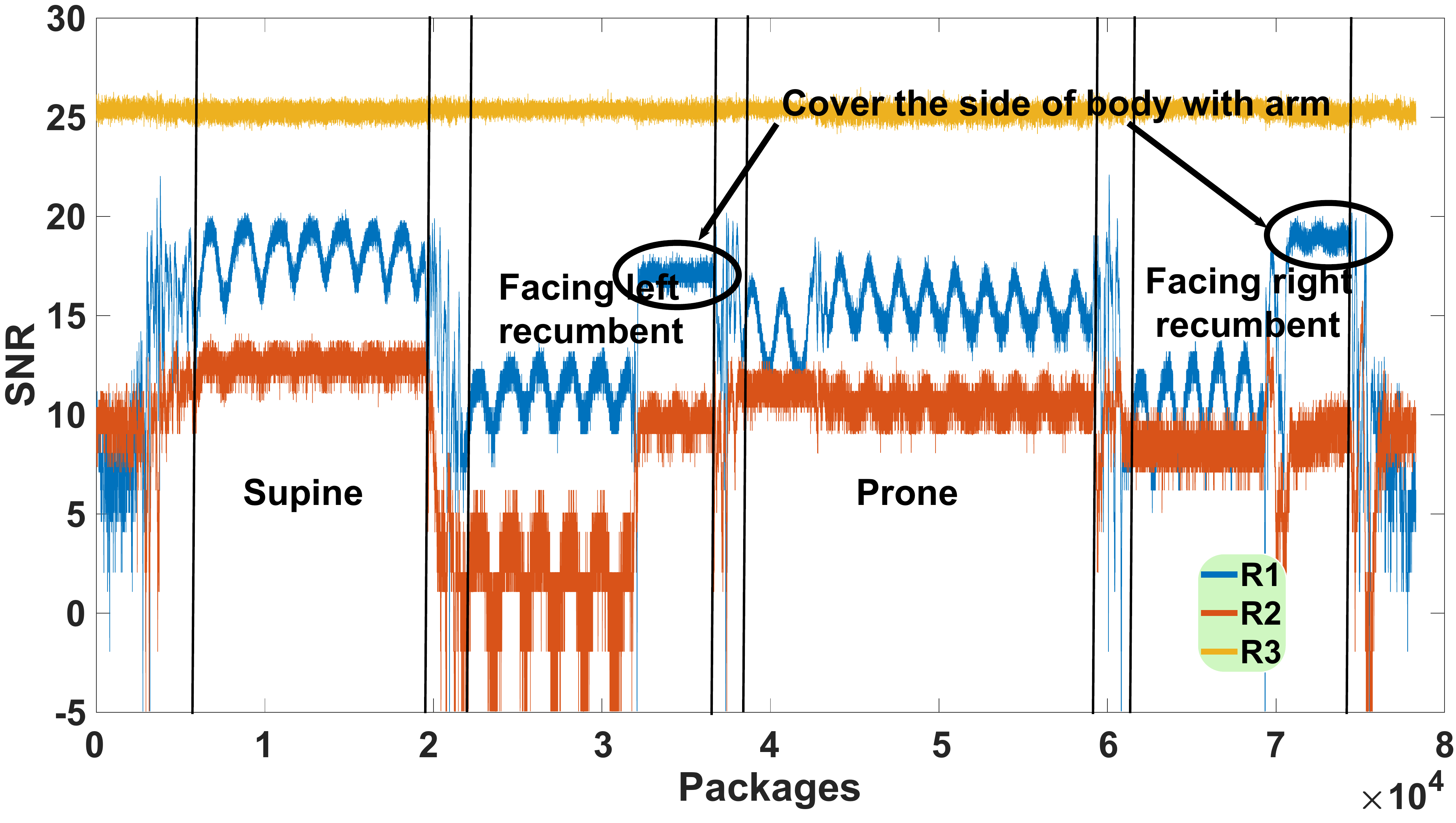}
\caption{Obstruction of the LOS signal from T1 to R1 in setting 1. The K factor of the T1--R1 stream decreases to 12.4, and the K factor of the T1--R3 stream increases to 5076.}
	\label{fig:plan1m}
\end{figure}

The performance of vital sign detection varies greatly among different people and different sleeping postures within the FFZ. Therefore, in our experiments, we constructed our prototype system in accordance with setting 1, and we selected the T1--R3 antenna pair to monitor vital signs. Due to the limitations imposed by the bed, the torso is closer to R3 when the human target is lying either face up or face down;
in other words, the torso is far from the mid-perpendicular of T1--R3. According to the Fresnel zone model, the chest displacement caused by breathing/heartbeats in the anteroposterior dimension still has a significant effect on T1--R3, thereby ensuring the monitoring performance in different sleeping postures. If we were to select the T1--R1 antenna pair, however, as shown in Fig. \ref{fig:plan1m}, the monitoring performance would become worse when the person's arm is blocking the side. This is because the main factor affecting the CSI received by R1 when the target is in the left-/right-facing recumbent position is the torso deformation in the lateral abdomen, and the arm sometimes blocks this area. This is why we selected T1--R3 for vital sign monitoring instead of T1--R1.

\begin{figure}[ht]
	\centering
	\includegraphics[width=1\columnwidth]{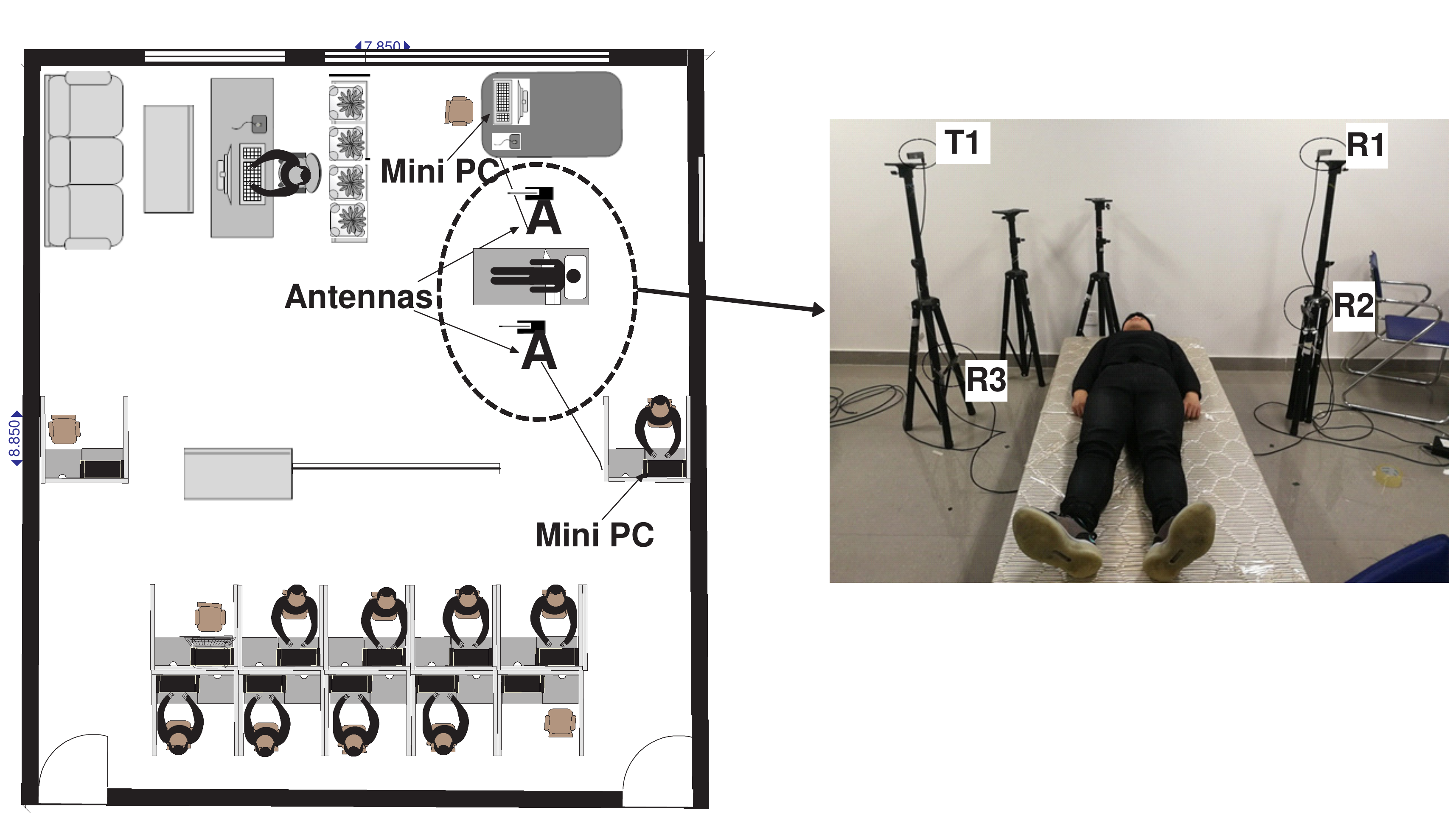}
\caption{The prototype system. T1, R1, R2, and R3 represent the transmit antenna and receive antennas 1, 2 and 3, respectively.}
	\label{fig:prototype}
\end{figure}

\begin{figure*}[htp]
	\label{se}
	\centering
\subfloat[]{\label{setup}
\begin{minipage}{0.3\linewidth}
			\centering
			\includegraphics[width=1\textwidth]{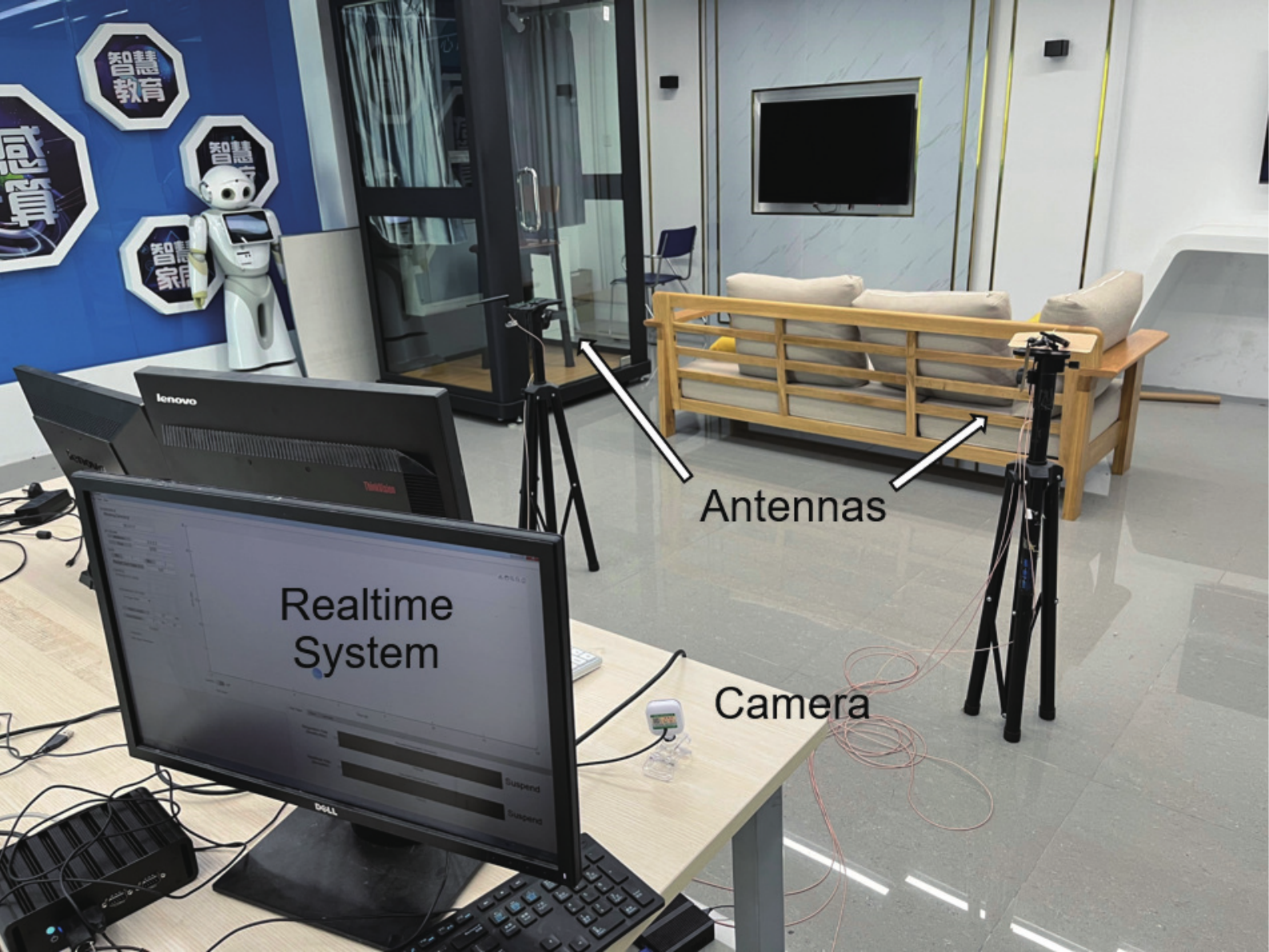}
\end{minipage}
}
\subfloat[]{\label{actvital}
\begin{minipage}{0.3\linewidth}
			\centering
			\includegraphics[width=1\textwidth]{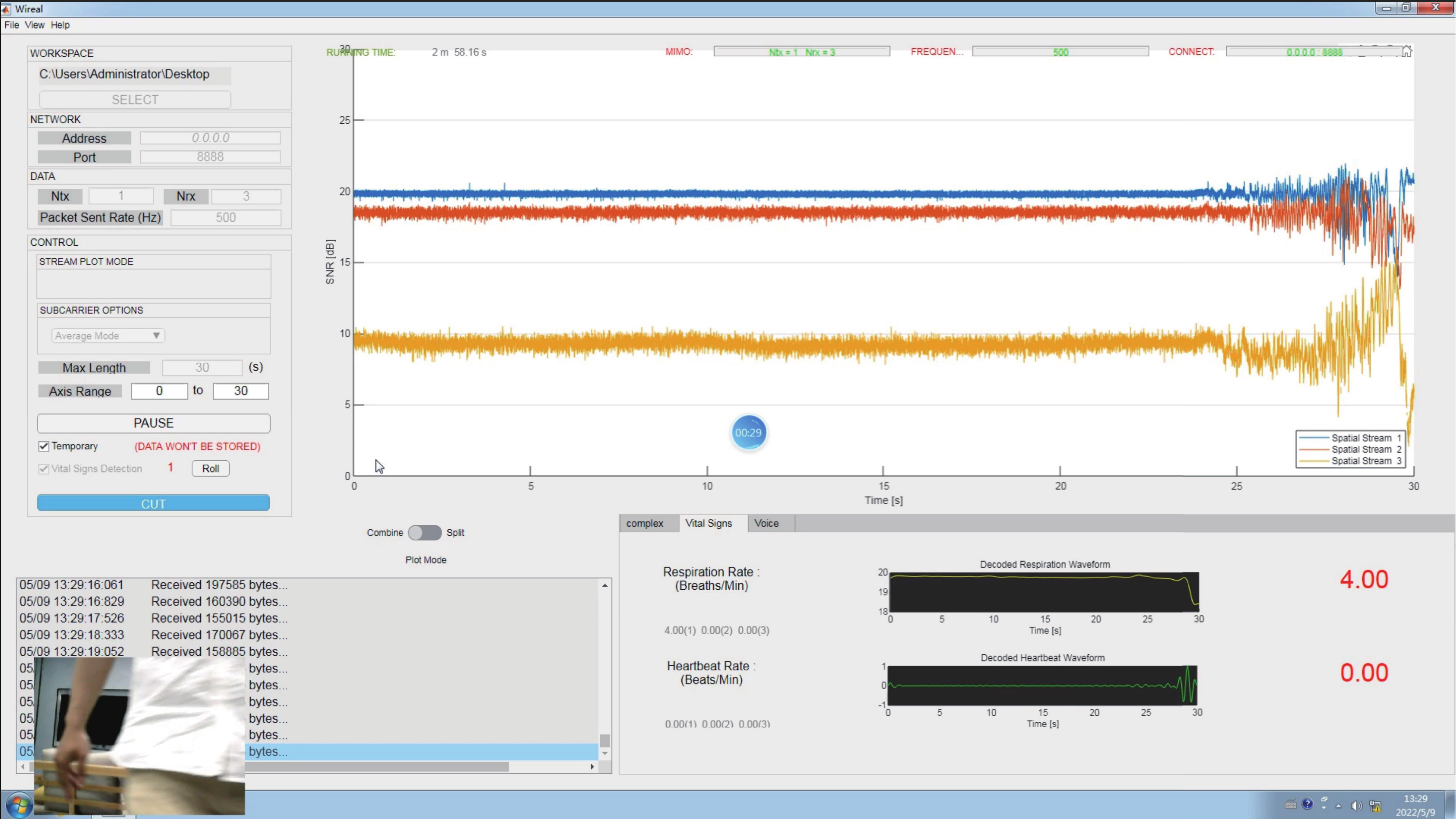}
\end{minipage}
}
\subfloat[]{\label{nomanvital}
\begin{minipage}{0.3\linewidth}
			\centering
			\includegraphics[width=1\textwidth]{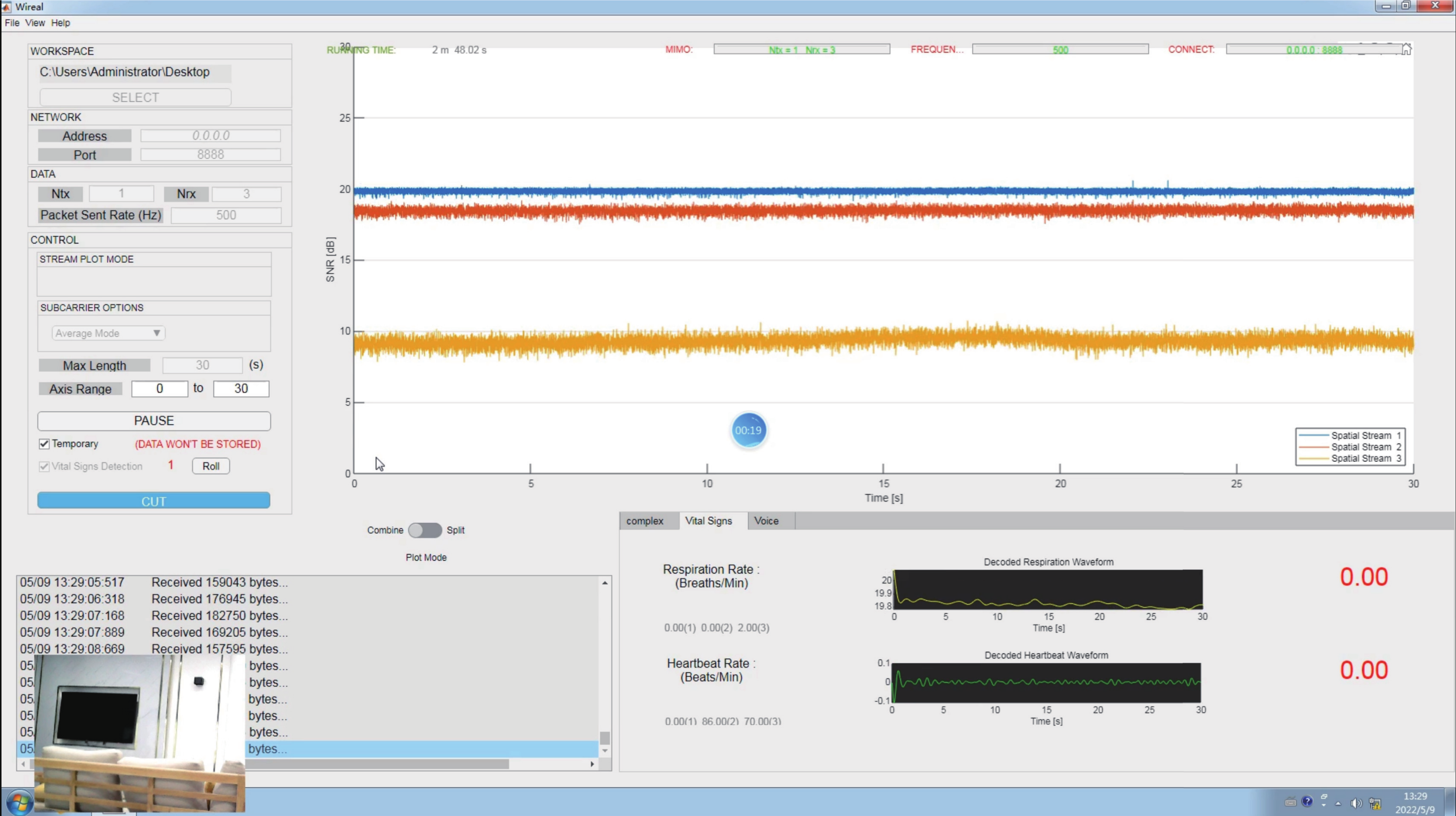}
\end{minipage}
}
\quad
\caption{System readings with no other person present and with active interference. (a) Setup. (b) With interference from other activities. (c) With no other person present.}
	\label{fig:noman}
\end{figure*}

In our experiments, we placed a lead plate under T1 to block the LOS path from T1 to R3 in accordance with our NLOS sensing model. CSI data were collected using csitool \cite{Halperin_csitool}, and the receiver transmitted the received CSI data to the monitoring computer for processing via the network.

We used COTS hardware devices to implement the proposed system. Specifically, we used two mini PCs with Intel Link 5300 Wi-Fi NICs as the transmitting and receiving devices. The mini PCs each had a 2.16 GHz Intel Celeron N2830 processor with 2 GB of RAM and Ubuntu OS version 12.04. The monitoring computer was a desktop computer equipped with an Intel Core i5 3450 CPU (3.1G Hz) and 2 GB of storage.


We conducted the experiments in a laboratory environment, as shown in Fig. \ref{fig:prototype}, with a total of 10 volunteers (6 male and 4 female) whose age range was 21 to 26 years. These 10 volunteers were university students who volunteered to participate in the experiments. During the experiments, we did not restrict the normal activities of others in the laboratory.

Each participant underwent a 30-minute actual test in different natural sleeping positions (prone, supine, left-facing recumbent, and right-facing recumbent). Different from previous work \cite{liu2018monitoring,zhang2019breathtrack}, we did not use a metronome to control the volunteers' respiratory rate, and we did not need to use a directional antenna to monitor the heart rate under LOS conditions. The ground-truth respiratory rate and heart rate were measured by an accelerometer attached to the abdomen and a fingertip pulse oximeter, respectively.

\subsection{Evaluation Results}

\begin{figure}[htp]
	\centering
	\includegraphics[width=0.8\columnwidth]{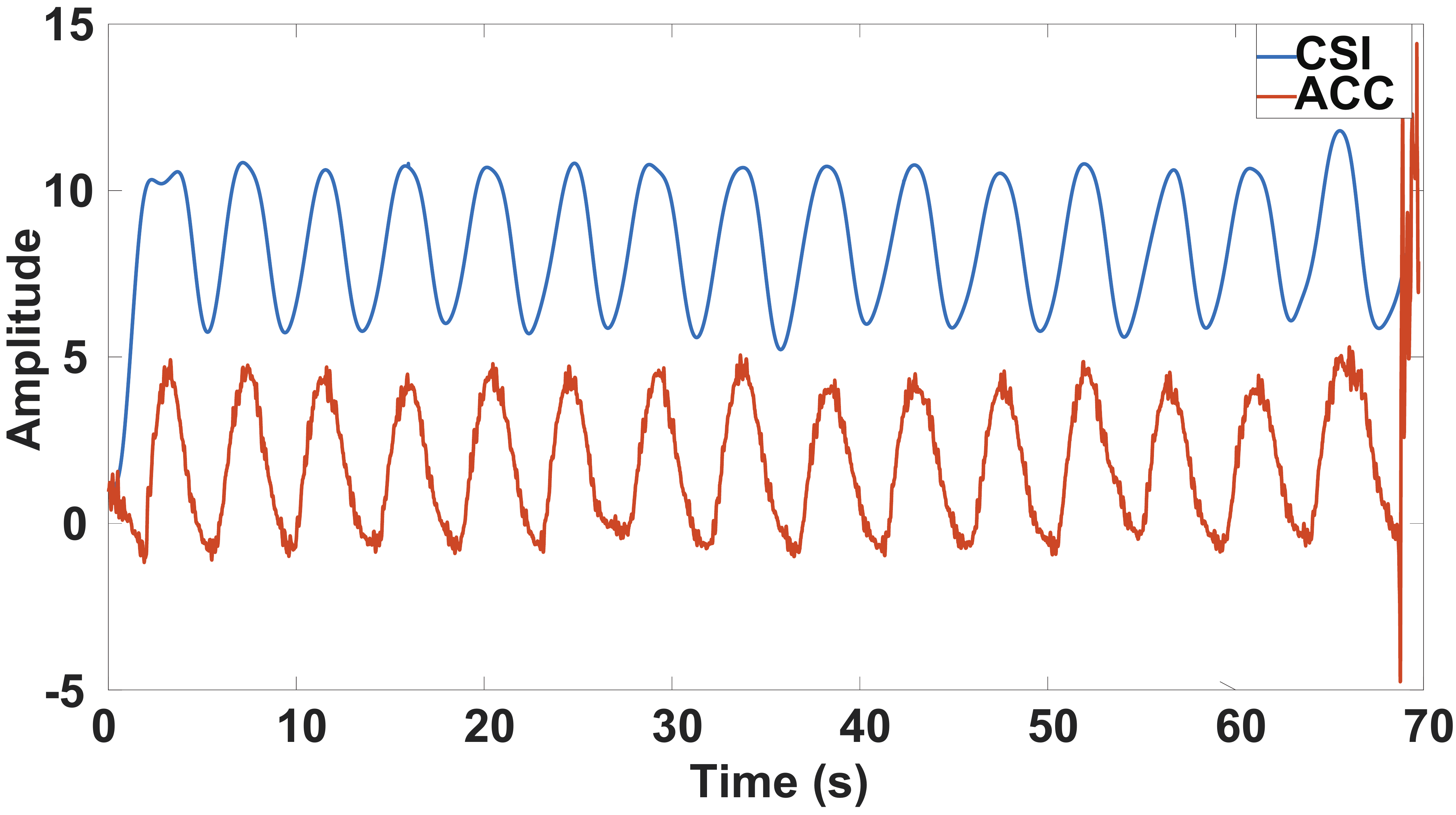}
\caption{Comparison of the processed CSI and the accelerometer (ACC) readings for breathing.}
	\label{fig:breath}
\end{figure}

\begin{figure}[htp]
 	\centering
    \includegraphics[width=0.95\columnwidth]{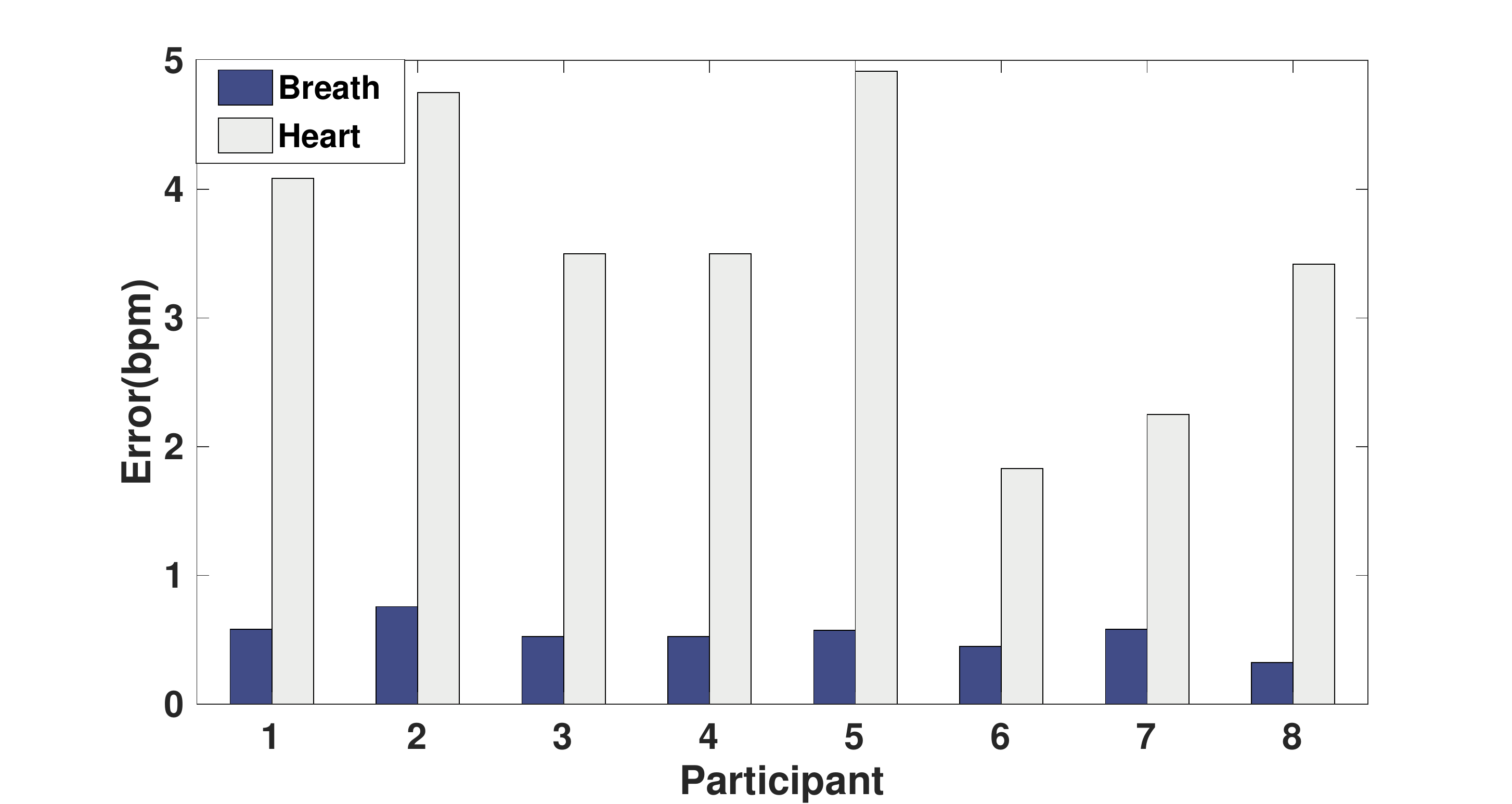}
\caption{Errors of vital sign estimation for different volunteers.}
    \label{fig:participants}
\end{figure}

Fig. \ref{fig:noman} shows the system readings in the presence of people performing other activities and in an unoccupied room. Our system does not output false respiratory/heart rate values in either of these scenarios.

\begin{figure}[htp]
	\centering
	\includegraphics[width=0.95\columnwidth]{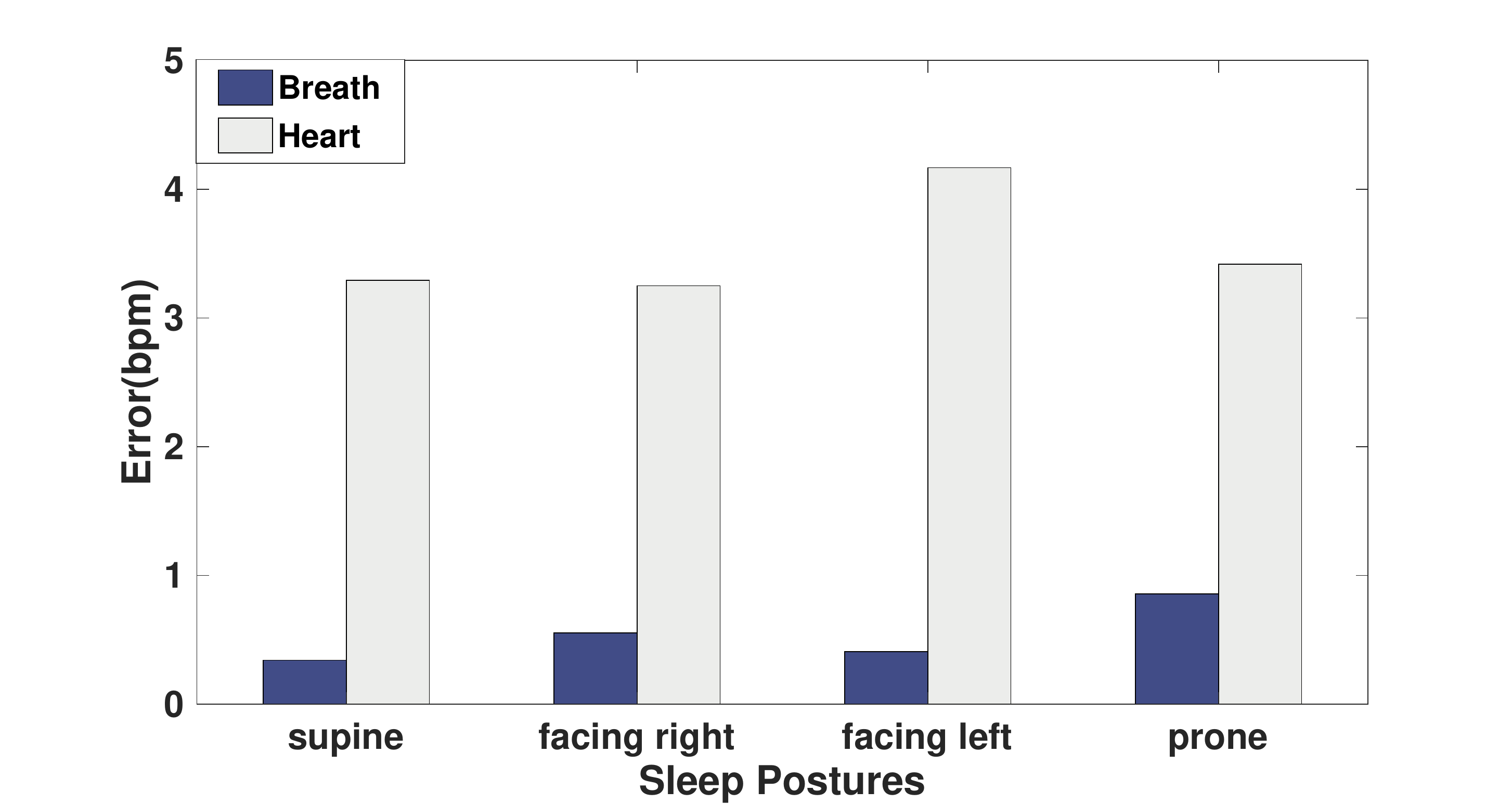}
\caption{Errors of vital sign estimation for different sleeping postures.}
	\label{fig:postures}
\end{figure}

We compared the obtained respiratory waveforms with the data from the accelerometer attached to the abdomen, as shown in Fig. \ref{fig:breath}. From this figure, we can observe that the CSI waveform is highly consistent with the respiratory waveform obtained by the accelerometer. Fig. \ref{fig:heart} compares heartbeat waveforms obtained in different postures with the readings from the accelerometer attached to the chest, and we find that the heartbeat readings obtained from the accelerometer are also consistent with the results of CSI detection. These results indicate that the CSI obtained from Wi-Fi signals can be used to extract fine-grained heartbeat and respiration information.

We evaluated the overall performance of breathing and heart rate estimation in different sleeping postures. The final results indicate average errors of 0.498 bpm (beats per minute) for the detected breathing rate and 3.531 bpm for the detected heart rate, and the corresponding accuracy is 96.887\% and 94.708\%, respectively.

Fig. \ref{fig:participants} illustrates the errors of vital sign (respiration and heart rate) monitoring for different participants; the volunteers had different body types, which led to different final results. However, in general, our system shows high accuracy in detecting respiration, and the error in detecting the heart rate is also within the acceptable range for nonclinical environments.

\begin{figure*}[ht]
	\label{RR}
	\centering
\subfloat[]{\label{R1}
\begin{minipage}{0.32\linewidth}
			\centering
			\includegraphics[width=1\textwidth]{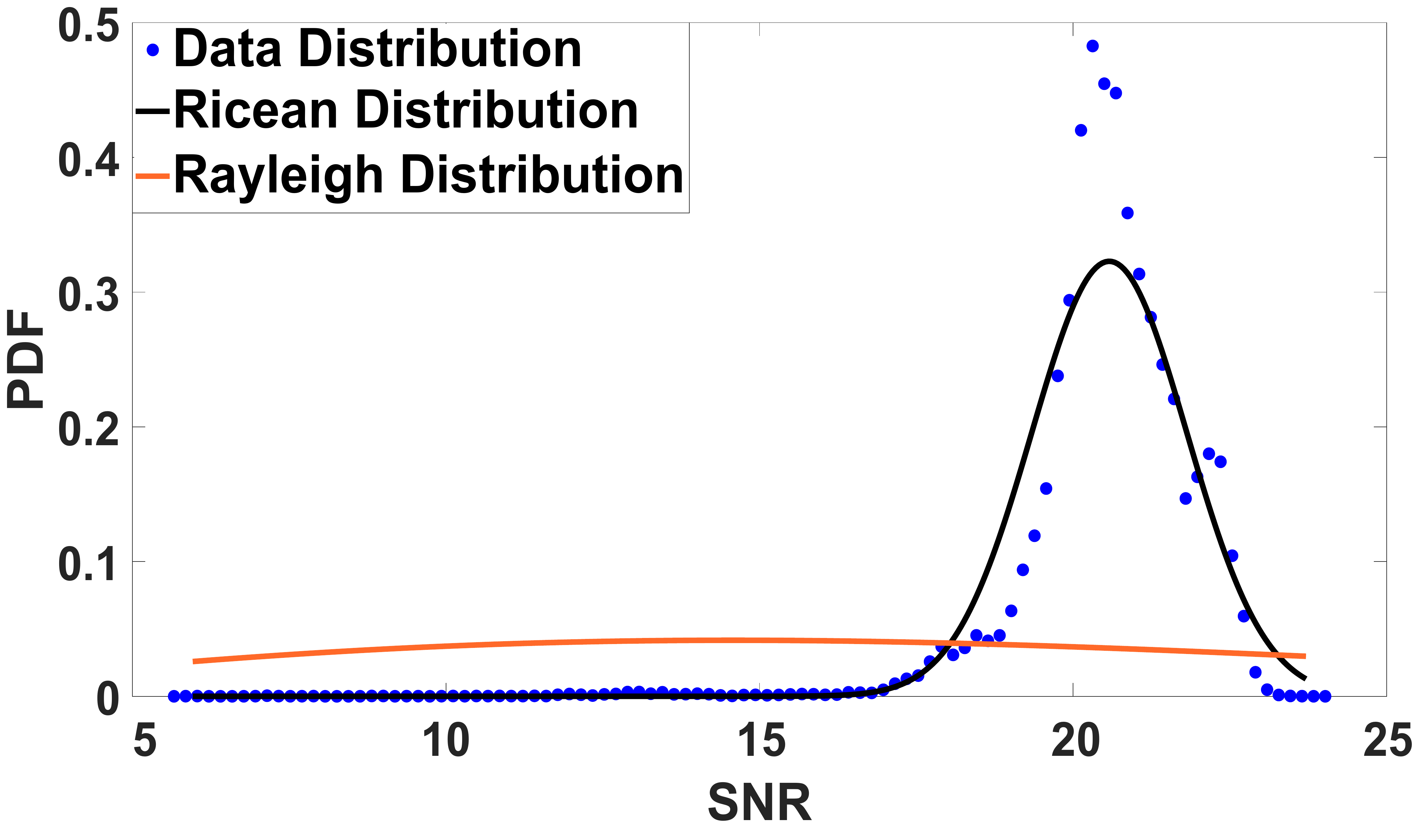}
\end{minipage}
}
\subfloat[]{\label{R2}
\begin{minipage}{0.32\linewidth}
			\centering
			\includegraphics[width=1\textwidth]{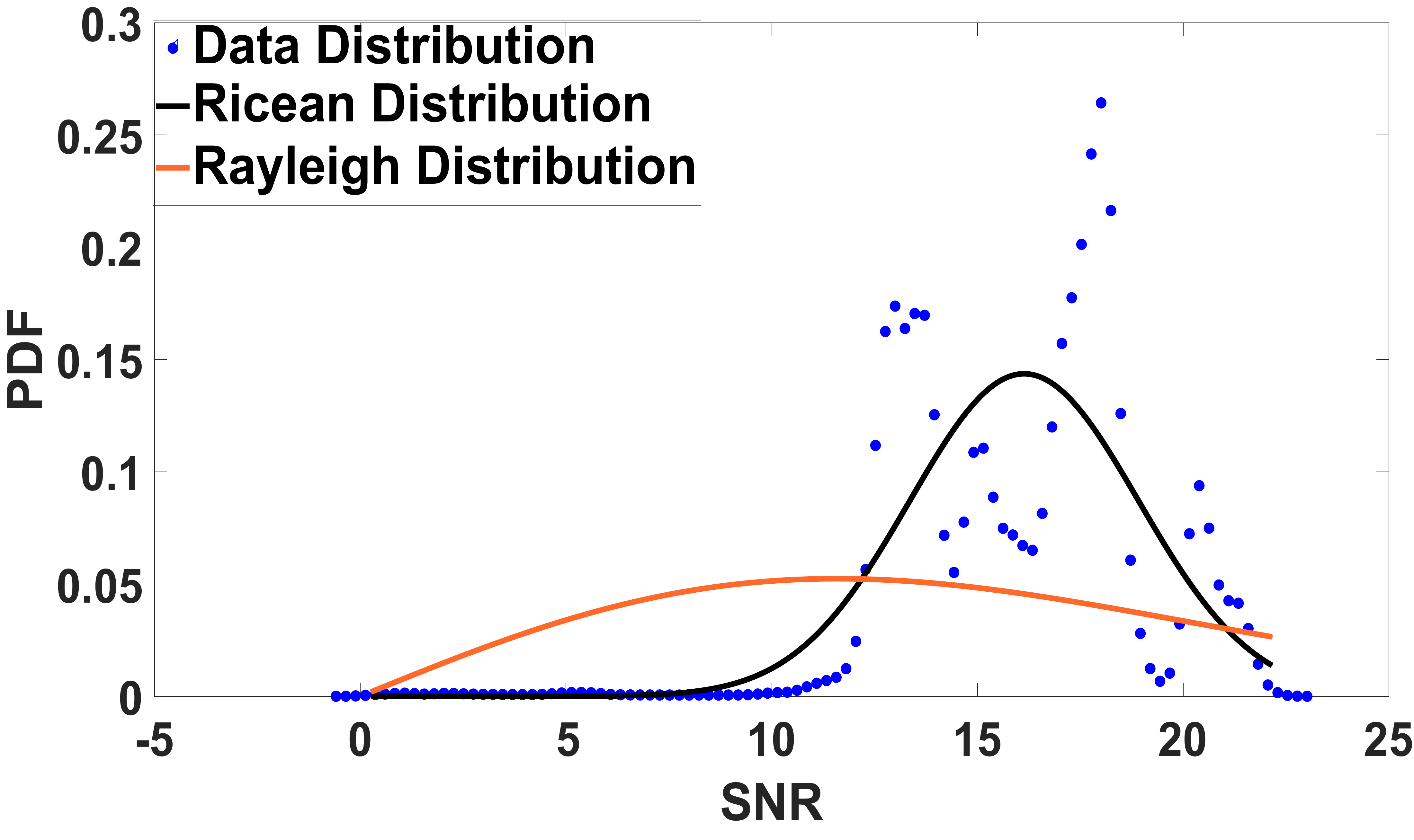}
\end{minipage}
}
\subfloat[]{\label{R3}
\begin{minipage}{0.32\linewidth}
			\centering
			\includegraphics[width=1\textwidth]{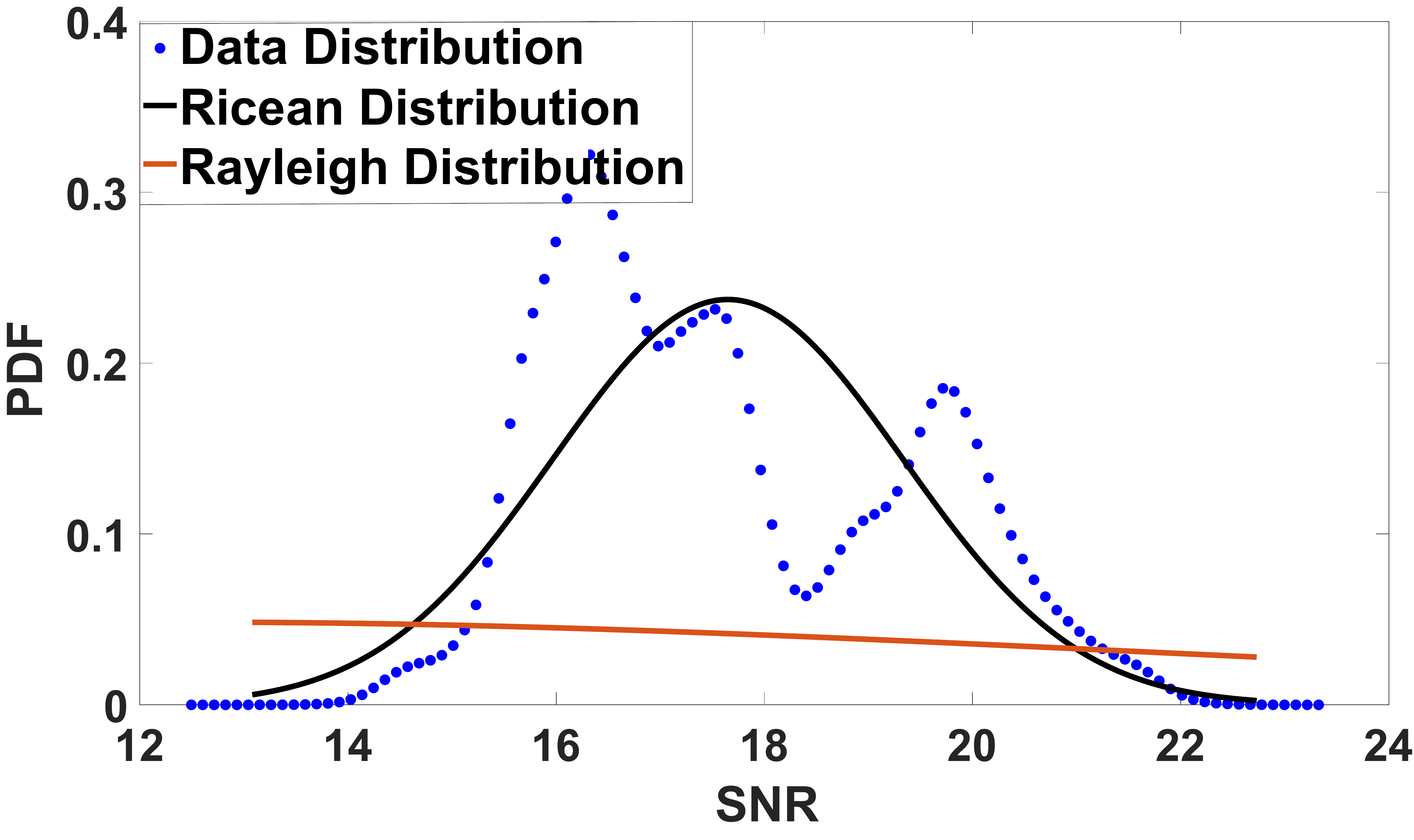}
\end{minipage}
}
\quad
	\caption{Distributions of the three streams in Fig.~\ref{plan1}. (a) T1--R1, $K=201.1$. (b) T1--R2, $K=17.8$. (c) T1--R3, $K=52$. PDF means probability density function.}
	\label{fig:KF}
\end{figure*}

Fig. \ref{fig:postures} illustrates the vital sign (respiration and heart rate) monitoring errors in different sleeping postures. These errors are relatively small in the supine and right-facing recumbent postures. However, the largest error in the monitored heart rate was found for the left-facing recumbent posture, while the largest error in the monitored respiratory rate was found for the prone posture. This is because the effective displacements of the dynamic path caused by these two sleeping postures are relatively small and not conducive to vital sign monitoring. Overall, our system can accurately monitor vital signs in different sleeping postures.

\subsection{Evaluation of the NLOS Sensing Model}

\begin{figure}[ht]
	\centering
	\includegraphics[width=0.9\columnwidth]{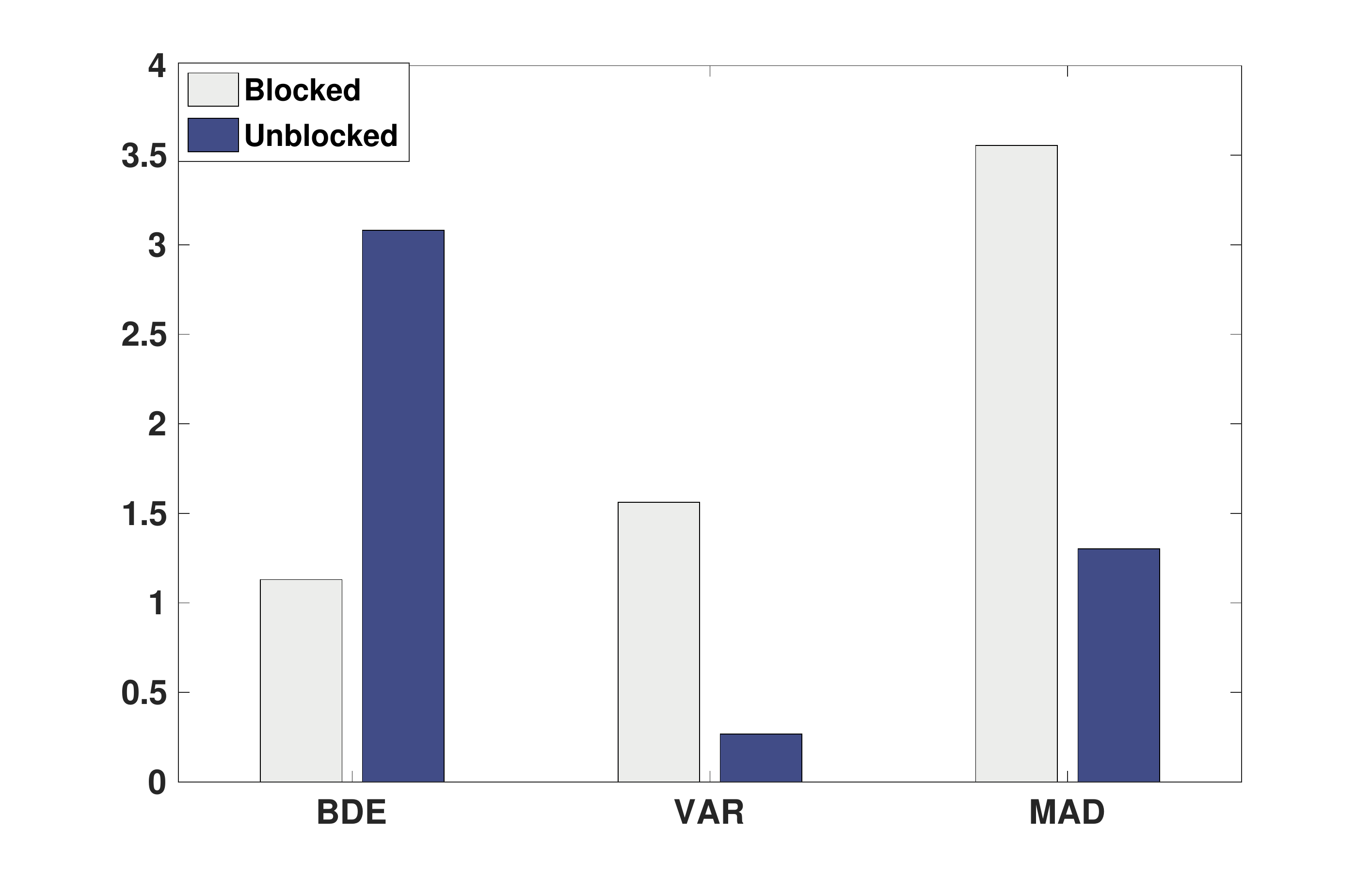}
\caption{The average breathing detection error (BDE), the variance of the CSI waveform (VAR), and the mean amplitude difference (MAD) when T1--R1 was blocked and unblocked.}
	\label{fig:RE}
\end{figure}

To verify the NLOS sensing model proposed in section \ref{Sect:pre}, we first calculated the Ricean K value of each stream in setting 1 (Fig. \ref{setting1}), as shown in Fig. \ref{fig:KF}, from which it can be found that the larger the value of $K$ is, the worse the sensing capability. Then, we placed a lead sheet between the T1--R1 antenna pair in setting 1 (decreasing $K$) and performed a breathing monitoring experiment, and the results are presented in Fig.~\ref{fig:plan1m}, showing that the motion sensing capability of T1--R1 was significantly improved. We also show the average breathing detection error (BDE), the variance of the CSI waveform (VAR), and the mean amplitude difference (MAD) in Fig. \ref{fig:RE}. We can observe improvements in both the detection accuracy and the sensitivity to motion.


\subsection{Comparison with Previous Approaches}
In this subsection, we compare Wital with four previously reported Wi-Fi-based vital sign monitoring approaches \cite{liu2015tracking,7980063,wang2017tensorbeat,9076681}.

\textbf{TVS.} Similar to our method, TVS \cite{liu2015tracking} uses the CSI amplitude for respiration sensing. First, the Hampel filter and a moving average filter are employed to clean the collected CSI data to remove noise. Then, subcarriers with larger variance are selected and subjected to the FFT to obtain the final results.

\textbf{PhaseBeat.} PhaseBeat \cite{7980063} uses the phase difference between the CSI readings of two antennas for respiration sensing. First, environment detection, data calibration, subcarrier selection and the discrete wavelet transform are used to preprocess the captured data. Then, peak detection is employed to estimate the respiratory rate.

\textbf{TensorBeat.} TensorBeat \cite{wang2017tensorbeat} is also a phase-difference-based method. First, the obtained CSI phase difference data are used to create tensors, and then, canonical polyadic (CP) decomposition is employed to obtain the respiration signals. Finally, peak detection is used to estimate the respiratory rate.

\textbf{ResBeat.} ResBeat \cite{9076681} uses both phase difference and amplitude for respiration monitoring. First, an adaptive signal selection module is leveraged to select the most sensitive signal group from the CSI amplitude groups and CSI phase difference groups. Then, peak detection is used to estimate the respiratory rate.

\begin{figure}[ht]
	\centering
	\includegraphics[width=0.9\columnwidth]{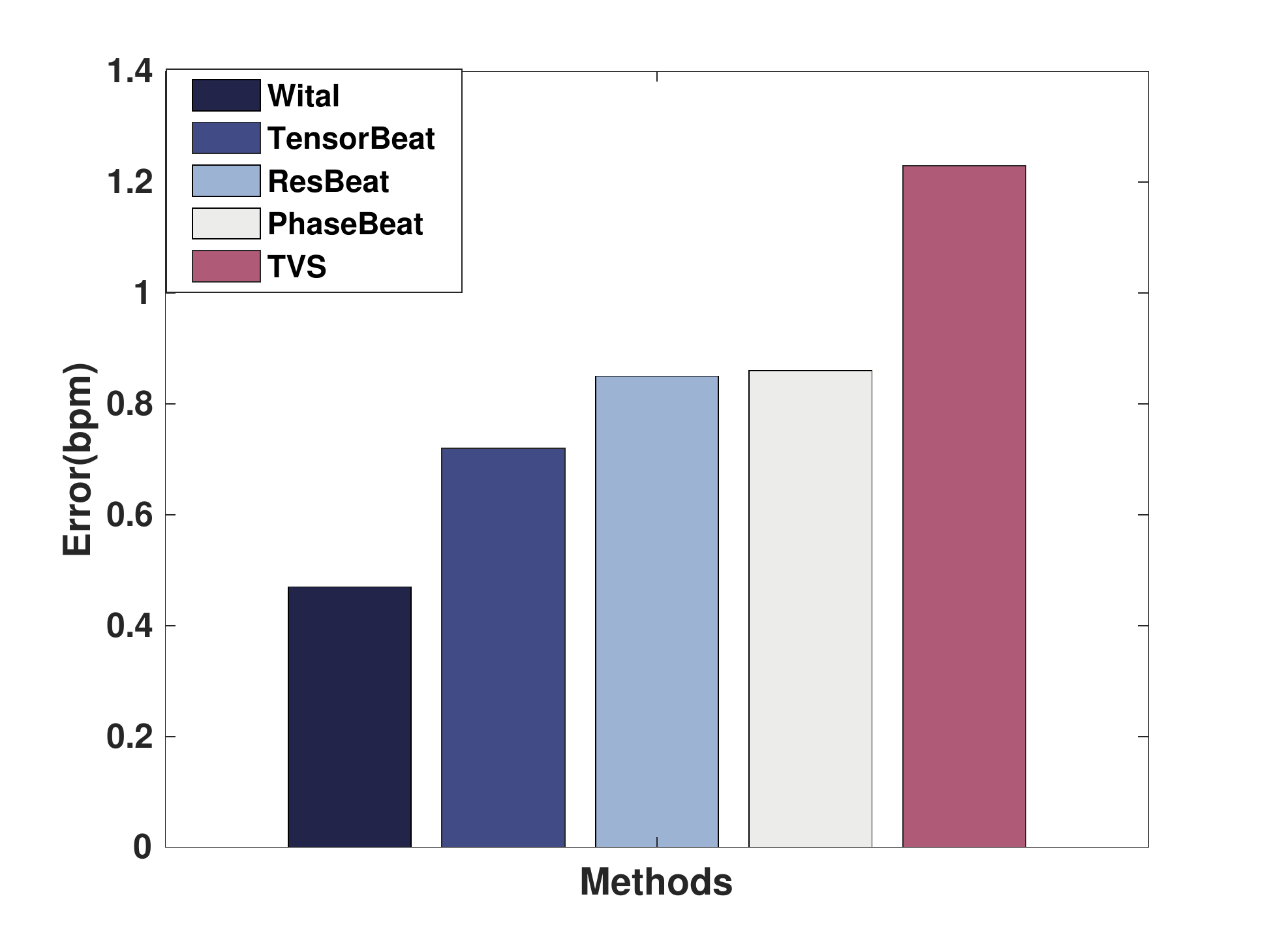}
\caption{Comparison with previous approaches.}
	\label{fig:comsota}
\end{figure}

Fig.~\ref{fig:comsota} presents the results. Wital achieves better performance than the other solutions. Although Wital's data processing is less sophisticated than that of the other solutions, we improve the respiration sensing ability by attenuating the LOS signal under the guidance of the proposed NLOS sensing model.

\subsection{Limitations and Future Work}

There are still some limitations to the application of our system in real-world scenarios, and these limitations are driving our further research. In this subsection, we describe these limitations and the solutions we propose to explore.

The first limitation is that Wital cannot recover the exact heartbeat waveform. Since the MATLAB-based system has difficulty handling multiple threads, we have used a crude bandpass filter and the FFT to obtain the heartbeats; however, this approach is not sufficient to obtain highly accurate and timely vital signs. The difficulty in overcoming this problem is that the trunk deformation caused by heartbeats can be overwhelmed by that caused by breathing, and it is difficult to separate heartbeats from breaths. The second limitation is that we are currently discarding the time periods during which interfering sleep activities occur, but it is also important to monitor vital signs during these periods. The challenge in solving this limitation is that such activities can overwhelm the vital signs. The third limitation is that for multiperson scenarios, we may be able to acquire the respiratory rates of multiple people (with different respiratory rates) via the FFT, but recovering the vital sign waveforms of individual people is difficult because of the superposition of the signal changes caused by multiple persons breathing.

Fortunately, these three problems can be viewed as the same general problem of separating signals that are superimposed on each other. Solutions based on traditional independent component analysis (ICA) may not solve these problems very well. However, we have reviewed some literature and believe that methods based on big data and deep learning may have unexpected benefits. For example, \cite{hyvarinen2017nonlinear} developed a nonlinear generalization approach based on temporal dependencies (e.g., autocorrelations), contrastive learning and a multilayer perceptron (MLP) for ICA or blind source separation. Some other literature on learning disentangled representations has focused on variational autoencoders (VAEs) \cite{locatello2019challenging, mita2021identifiable}. The authors of these studies first encode the original input to obtain latent variables and impose artificially specified constraints on these latent variables to force the approximate posterior to match the prior in the latent space (under the assumption that the latent variables obtained by the VAE approximate the true latent variables of the input). Then, they reconstruct the input using a decoder and the obtained latent variables. Finally, a well-trained VAE can disentangle the factors (latent variables) of interest from the inputs. The core of both approaches is to find a suitable and effective model structure and constraints; these constraints should reflect the differences in the factors that are to be separated. We are currently experimenting with these solutions to enhance the applicability of our system.

\section{Conclusion}
\label{Sect:con}
In this paper, we show the ability to use Wi-Fi signals to track breathing and heartbeats in different sleeping postures using only one pair of Wi-Fi devices. To achieve this, we propose an NLOS sensing model based on Ricean K theory to help monitor the minor displacements caused by breathing and heartbeats, and we theoretically prove that blocking the LOS signal is beneficial for motion detection in NLOS sensing. We also propose a motion segmentation method based on regularity detection, which can accurately identify the time ranges corresponding to motions different from vital signs (such as turning over and rising from bed). We have implemented a prototype system to evaluate our method. The experimental results indicate that our method achieves 96.618\% and 94.708\% accuracy for breathing and heart rate detection, respectively.

\bibliographystyle{IEEEtran}
\bibliography{Vital}

\begin{thebibliography}{10}
\providecommand{\url}[1]{#1}
\csname url@samestyle\endcsname
\providecommand{\newblock}{\relax}
\providecommand{\bibinfo}[2]{#2}
\providecommand{\BIBentrySTDinterwordspacing}{\spaceskip=0pt\relax}
\providecommand{\BIBentryALTinterwordstretchfactor}{4}
\providecommand{\BIBentryALTinterwordspacing}{\spaceskip=\fontdimen2\font plus
\BIBentryALTinterwordstretchfactor\fontdimen3\font minus
  \fontdimen4\font\relax}
\providecommand{\BIBforeignlanguage}[2]{{%
\expandafter\ifx\csname l@#1\endcsname\relax
\typeout{** WARNING: IEEEtran.bst: No hyphenation pattern has been}%
\typeout{** loaded for the language `#1'. Using the pattern for}%
\typeout{** the default language instead.}%
\else
\language=\csname l@#1\endcsname
\fi
#2}}
\providecommand{\BIBdecl}{\relax}
\BIBdecl

\bibitem{min2010noncontact}
S.~D. Min, J.~K. Kim, H.~S. Shin, Y.~H. Yun, C.~K. Lee, and M.~Lee,
  ``Noncontact respiration rate measurement system using an ultrasonic
  proximity sensor,'' \emph{IEEE Sensors Journal}, vol.~10, no.~11, pp.
  1732--1739, 2010.

\bibitem{braun2012bridging}
P.~X. Braun, C.~F. Gmachl, and R.~A. Dweik, ``Bridging the collaborative gap:
  Realizing the clinical potential of breath analysis for disease diagnosis and
  monitoring--tutorial,'' \emph{IEEE Sensors Journal}, vol.~12, no.~11, pp.
  3258--3270, 2012.

\bibitem{facco2014sleep}
F.~L. Facco, D.~W. Ouyang, P.~C. Zee, and W.~A. Grobman, ``Sleep disordered
  breathing in a high-risk cohort prevalence and severity across pregnancy,''
  \emph{American journal of perinatology}, vol.~31, no.~10, pp. 899--904, 2014.

\bibitem{kushida2005practice}
C.~A. Kushida, M.~R. Littner, T.~Morgenthaler, C.~A. Alessi, D.~Bailey,
  J.~Coleman~Jr, L.~Friedman, M.~Hirshkowitz, S.~Kapen, M.~Kramer
  \emph{et~al.}, ``Practice parameters for the indications for polysomnography
  and related procedures: an update for 2005,'' \emph{Sleep}, vol.~28, no.~4,
  pp. 499--523, 2005.

\bibitem{nappholz1992implantable}
T.~A. Nappholz, W.~N. Hursta, A.~K. Dawson, and B.~M. Steinhaus, ``Implantable
  ambulatory electrocardiogram monitor,'' May~19 1992, uS Patent 5,113,869.

\bibitem{zhao2016emotion}
M.~Zhao, F.~Adib, and D.~Katabi, ``Emotion recognition using wireless
  signals,'' in \emph{Proceedings of the 22nd Annual International Conference
  on Mobile Computing and Networking}, 2016, pp. 95--108.

\bibitem{yue2018extracting}
S.~Yue, H.~He, H.~Wang, H.~Rahul, and D.~Katabi, ``Extracting multi-person
  respiration from entangled rf signals,'' \emph{Proceedings of the ACM on
  Interactive, Mobile, Wearable and Ubiquitous Technologies}, vol.~2, no.~2,
  pp. 1--22, 2018.

\bibitem{zhang2018fresnel}
F.~Zhang, D.~Zhang, J.~Xiong, H.~Wang, K.~Niu, B.~Jin, and Y.~Wang, ``From
  fresnel diffraction model to fine-grained human respiration sensing with
  commodity wi-fi devices,'' \emph{Proceedings of the ACM on Interactive,
  Mobile, Wearable and Ubiquitous Technologies}, vol.~2, no.~1, p.~53, 2018.

\bibitem{zeng2019farsense}
Y.~Zeng, D.~Wu, J.~Xiong, E.~Yi, R.~Gao, and D.~Zhang, ``Farsense: Pushing the
  range limit of wifi-based respiration sensing with csi ratio of two
  antennas,'' \emph{Proceedings of the ACM on Interactive, Mobile, Wearable and
  Ubiquitous Technologies}, vol.~3, no.~3, pp. 1--26, 2019.

\bibitem{9076681}
X.~Wang, C.~Yang, and S.~Mao, ``Resilient respiration rate monitoring with
  realtime bimodal csi data,'' \emph{IEEE Sensors Journal}, vol.~20, no.~17,
  pp. 10\,187--10\,198, 2020.

\bibitem{liu2018monitoring}
J.~Liu, Y.~Chen, Y.~Wang, X.~Chen, J.~Cheng, and J.~Yang, ``Monitoring vital
  signs and postures during sleep using wifi signals,'' \emph{IEEE Internet of
  Things Journal}, vol.~5, no.~3, pp. 2071--2084, 2018.

\bibitem{wang2016human}
H.~Wang, D.~Zhang, J.~Ma, Y.~Wang, Y.~Wang, D.~Wu, T.~Gu, and B.~Xie, ``Human
  respiration detection with commodity wifi devices: do user location and body
  orientation matter?'' in \emph{Proceedings of the 2016 ACM International
  Joint Conference on Pervasive and Ubiquitous Computing}.\hskip 1em plus 0.5em
  minus 0.4em\relax ACM, 2016, pp. 25--36.

\bibitem{zeng2020multisense}
Y.~Zeng, D.~Wu, J.~Xiong, J.~Liu, Z.~Liu, and D.~Zhang, ``Multisense: Enabling
  multi-person respiration sensing with commodity wifi,'' \emph{Proceedings of
  the ACM on Interactive, Mobile, Wearable and Ubiquitous Technologies},
  vol.~4, no.~3, pp. 1--29, 2020.

\bibitem{8047246}
Y.~Gu, J.~Zhan, Y.~Ji, J.~Li, F.~Ren, and S.~Gao, ``Mosense: An rf-based motion
  detection system via off-the-shelf wifi devices,'' \emph{IEEE Internet of
  Things Journal}, vol.~4, no.~6, pp. 2326--2341, 2017.

\bibitem{liu2016contactless}
X.~Liu, J.~Cao, S.~Tang, J.~Wen, and P.~Guo, ``Contactless respiration
  monitoring via off-the-shelf wifi devices,'' \emph{IEEE Transactions on
  Mobile Computing}, vol.~15, no.~10, pp. 2466--2479, 2016.

\bibitem{zhang2019breathtrack}
D.~Zhang, Y.~Hu, Y.~Chen, and B.~Zeng, ``Breathtrack: Tracking indoor human
  breath status via commodity wifi,'' \emph{IEEE Internet of Things Journal},
  vol.~6, no.~2, pp. 3899--3911, 2019.

\bibitem{liu2021wiphone}
J.~Liu, Y.~Zeng, T.~Gu, L.~Wang, and D.~Zhang, ``Wiphone: Smartphone-based
  respiration monitoring using ambient reflected wifi signals,''
  \emph{Proceedings of the ACM on Interactive, Mobile, Wearable and Ubiquitous
  Technologies}, vol.~5, no.~1, pp. 1--19, 2021.

\bibitem{ma2019wifi}
Y.~Ma, G.~Zhou, and S.~Wang, ``Wifi sensing with channel state information: A
  survey,'' \emph{ACM Computing Surveys (CSUR)}, vol.~52, no.~3, pp. 1--36,
  2019.

\bibitem{gu2019besense}
Y.~Gu, X.~Zhang, Z.~Liu, and F.~Ren, ``Besense: Leveraging wifi channel data
  and computational intelligence for behavior analysis,'' \emph{IEEE
  Computational Intelligence Magazine}, vol.~14, no.~4, pp. 31--41, 2019.

\bibitem{huang2020towards}
J.~Huang, B.~Liu, P.~Liu, C.~Chen, N.~Xiao, Y.~Wu, C.~Zhang, and N.~Yu,
  ``Towards anti-interference wifi-based activity recognition system using
  interference-independent phase component,'' in \emph{IEEE INFOCOM 2020-IEEE
  Conference on Computer Communications}.\hskip 1em plus 0.5em minus
  0.4em\relax IEEE, 2020, pp. 576--585.

\bibitem{huang2021phaseanti}
J.~Huang, B.~Liu, C.~Miao, Y.~Lu, Q.~Zheng, Y.~Wu, J.~Liu, L.~Su, and C.~W.
  Chen, ``Phaseanti: an anti-interference wifi-based activity recognition
  system using interference-independent phase component,'' \emph{IEEE
  Transactions on Mobile Computing}, pp. 1--1, 2021.

\bibitem{9075376}
J.~Huang, B.~Liu, C.~Chen, H.~Jin, Z.~Liu, C.~Zhang, and N.~Yu, ``Towards
  anti-interference human activity recognition based on wifi subcarrier
  correlation selection,'' \emph{IEEE Transactions on Vehicular Technology},
  vol.~69, no.~6, pp. 6739--6754, 2020.

\bibitem{huang2020widet}
H.~Huang and S.~Lin, ``Widet: Wi-fi based device-free passive person detection
  with deep convolutional neural networks,'' \emph{Computer Communications},
  vol. 150, pp. 357--366, 2020.

\bibitem{ahmed2020device}
H.~F.~T. Ahmed, H.~Ahmad, and C.~Aravind, ``Device free human gesture
  recognition using wi-fi csi: A survey,'' \emph{Engineering Applications of
  Artificial Intelligence}, vol.~87, p. 103281, 2020.

\bibitem{qian2017widar}
K.~Qian, C.~Wu, Z.~Yang, Y.~Liu, and K.~Jamieson, ``Widar: Decimeter-level
  passive tracking via velocity monitoring with commodity wi-fi,'' in
  \emph{Proceedings of the 18th ACM International Symposium on Mobile Ad Hoc
  Networking and Computing}, 2017, pp. 1--10.

\bibitem{8820006}
H.~Yan, Y.~Zhang, Y.~Wang, and K.~Xu, ``Wiact: A passive wifi-based human
  activity recognition system,'' \emph{IEEE Sensors Journal}, vol.~20, no.~1,
  pp. 296--305, 2020.

\bibitem{zhang2021widar3}
Y.~Zhang, Y.~Zheng, K.~Qian, G.~Zhang, Y.~Liu, C.~Wu, and Z.~Yang, ``Widar3.0:
  Zero-effort cross-domain gesture recognition with wi-fi,'' \emph{IEEE
  Transactions on Pattern Analysis and Machine Intelligence}, vol.~44, no.~11,
  pp. 8671--8688, 2022.

\bibitem{gu2022wigrunt}
Y.~Gu, X.~Zhang, Y.~Wang, M.~Wang, H.~Yan, Y.~Ji, Z.~Liu, J.~Li, and M.~Dong,
  ``Wigrunt: Wifi-enabled gesture recognition using dual-attention network,''
  \emph{IEEE Transactions on Human-Machine Systems}, vol.~52, no.~4, pp.
  736--746, 2022.

\bibitem{zhao2019accurate}
L.~Zhao, H.~Huang, X.~Li, S.~Ding, H.~Zhao, and Z.~Han, ``An accurate and
  robust approach of device-free localization with convolutional autoencoder,''
  \emph{IEEE Internet of Things Journal}, vol.~6, no.~3, pp. 5825--5840, 2019.

\bibitem{cao2019contactless}
Y.~Cao, F.~Wang, X.~Lu, N.~Lin, B.~Zhang, Z.~Liu, and S.~Sigg, ``Contactless
  body movement recognition during sleep via wifi signals,'' \emph{IEEE
  Internet of Things Journal}, vol.~7, no.~3, pp. 2028--2037, 2020.

\bibitem{bai2019widrive}
Y.~Bai, Z.~Wang, K.~Zheng, X.~Wang, and J.~Wang, ``Widrive: Adaptive wifi-based
  recognition of driver activity for real-time and safe takeover,'' in
  \emph{2019 IEEE 39th International Conference on Distributed Computing
  Systems (ICDCS)}.\hskip 1em plus 0.5em minus 0.4em\relax IEEE, 2019, pp.
  901--911.

\bibitem{wang2019person}
F.~Wang, S.~Zhou, S.~Panev, J.~Han, and D.~Huang, ``Person-in-wifi:
  Fine-grained person perception using wifi,'' in \emph{Proceedings of the IEEE
  International Conference on Computer Vision}, 2019, pp. 5452--5461.

\bibitem{wu2020fingerdraw}
D.~Wu, R.~Gao, Y.~Zeng, J.~Liu, L.~Wang, T.~Gu, and D.~Zhang, ``Fingerdraw:
  Sub-wavelength level finger motion tracking with wifi signals,''
  \emph{Proceedings of the ACM on Interactive, Mobile, Wearable and Ubiquitous
  Technologies}, vol.~4, no.~1, pp. 1--27, 2020.

\bibitem{meng2019revealing}
Y.~Meng, J.~Li, H.~Zhu, X.~Liang, Y.~Liu, and N.~Ruan, ``Revealing your mobile
  password via wifi signals: Attacks and countermeasures,'' \emph{IEEE
  Transactions on Mobile Computing}, vol.~19, no.~2, pp. 432--449, 2020.

\bibitem{zhang2019smars}
F.~Zhang, C.~Wu, B.~Wang, M.~Wu, D.~Bugos, H.~Zhang, and K.~R. Liu, ``Smars:
  Sleep monitoring via ambient radio signals,'' \emph{IEEE Transactions on
  Mobile Computing}, vol.~20, no.~1, pp. 217--231, 2019.

\bibitem{aly2016zephyr}
H.~Aly and M.~Youssef, ``Zephyr: Ubiquitous accurate multi-sensor fusion-based
  respiratory rate estimation using smartphones,'' in \emph{IEEE INFOCOM
  2016-The 35th Annual IEEE International Conference on Computer
  Communications}.\hskip 1em plus 0.5em minus 0.4em\relax IEEE, 2016, pp. 1--9.

\bibitem{paalasmaa2012unobtrusive}
J.~Paalasmaa, M.~Waris, H.~Toivonen, L.~Lepp{\"a}korpi, and M.~Partinen,
  ``Unobtrusive online monitoring of sleep at home,'' in \emph{2012 Annual
  International Conference of the IEEE Engineering in Medicine and Biology
  Society}.\hskip 1em plus 0.5em minus 0.4em\relax IEEE, 2012, pp. 3784--3788.

\bibitem{kumar2015distanceppg}
M.~Kumar, A.~Veeraraghavan, and A.~Sabharwal, ``Distanceppg: Robust non-contact
  vital signs monitoring using a camera,'' \emph{Biomedical optics express},
  vol.~6, no.~5, pp. 1565--1588, 2015.

\bibitem{salmi2011propagation}
J.~Salmi and A.~F. Molisch, ``Propagation parameter estimation, modeling and
  measurements for ultrawideband mimo radar,'' \emph{IEEE Transactions on
  Antennas and Propagation}, vol.~59, no.~11, pp. 4257--4267, 2011.

\bibitem{patwari2013breathfinding}
N.~Patwari, L.~Brewer, Q.~Tate, O.~Kaltiokallio, and M.~Bocca, ``Breathfinding:
  A wireless network that monitors and locates breathing in a home,''
  \emph{IEEE Journal of Selected Topics in Signal Processing}, vol.~8, no.~1,
  pp. 30--42, 2013.

\bibitem{patwari2013monitoring}
N.~Patwari, J.~Wilson, S.~Ananthanarayanan, S.~K. Kasera, and D.~R. Westenskow,
  ``Monitoring breathing via signal strength in wireless networks,'' \emph{IEEE
  Transactions on Mobile Computing}, vol.~13, no.~8, pp. 1774--1786, 2013.

\bibitem{abdelnasser2015ubibreathe}
H.~Abdelnasser, K.~A. Harras, and M.~Youssef, ``Ubibreathe: A ubiquitous
  non-invasive wifi-based breathing estimator,'' in \emph{Proceedings of the
  16th ACM International Symposium on Mobile Ad Hoc Networking and
  Computing}.\hskip 1em plus 0.5em minus 0.4em\relax ACM, 2015, pp. 277--286.

\bibitem{wu2015non}
C.~Wu, Z.~Yang, Z.~Zhou, X.~Liu, Y.~Liu, and J.~Cao, ``Non-invasive detection
  of moving and stationary human with wifi,'' \emph{IEEE Journal on Selected
  Areas in Communications}, vol.~33, no.~11, pp. 2329--2342, 2015.

\bibitem{chen2017tr}
C.~Chen, Y.~Han, Y.~Chen, H.-Q. Lai, F.~Zhang, B.~Wang, and K.~R. Liu,
  ``Tr-breath: Time-reversal breathing rate estimation and detection,''
  \emph{IEEE Transactions on Biomedical Engineering}, vol.~65, no.~3, pp.
  489--501, 2017.

\bibitem{wang2017tensorbeat}
X.~Wang, C.~Yang, and S.~Mao, ``Tensorbeat: Tensor decomposition for monitoring
  multiperson breathing beats with commodity wifi,'' \emph{ACM Transactions on
  Intelligent Systems and Technology (TIST)}, vol.~9, no.~1, pp. 1--27, 2017.

\bibitem{gu2021real}
Y.~Gu, X.~Zhang, H.~Yan, Z.~Liu, and Y.~Ji, ``Real-time vital signs monitoring
  based on cots wifi devices,'' in \emph{2021 IEEE International Conference on
  Bioinformatics and Biomedicine (BIBM)}.\hskip 1em plus 0.5em minus
  0.4em\relax IEEE, 2021, pp. 1320--1324.

\bibitem{9385792}
Y.~Gu, H.~Yan, M.~Dong, M.~Wang, X.~Zhang, Z.~Liu, and F.~Ren, ``Wione:
  One-shot learning for environment-robust device-free user authentication via
  commodity wi-fi in man–machine system,'' \emph{IEEE Transactions on
  Computational Social Systems}, vol.~8, no.~3, pp. 630--642, 2021.

\bibitem{zhang2019towards}
F.~Zhang, K.~Niu, J.~Xiong, B.~Jin, T.~Gu, Y.~Jiang, and D.~Zhang, ``Towards a
  diffraction-based sensing approach on human activity recognition,''
  \emph{Proceedings of the ACM on Interactive, Mobile, Wearable and Ubiquitous
  Technologies}, vol.~3, no.~1, pp. 1--25, 2019.

\bibitem{tepedelenlioglu2003ricean}
C.~Tepedelenlioglu, A.~Abdi, and G.~B. Giannakis, ``The ricean k factor:
  estimation and performance analysis,'' \emph{IEEE Transactions on Wireless
  Communications}, vol.~2, no.~4, pp. 799--810, 2003.

\bibitem{gu2019wi}
Y.~Gu, X.~Zhang, Z.~Liu, and F.~Ren, ``Wifi-based real-time breathing and heart
  rate monitoring during sleep,'' in \emph{2019 IEEE Global Communications
  Conference (GLOBECOM)}.\hskip 1em plus 0.5em minus 0.4em\relax IEEE, 2019,
  pp. 1--6.

\bibitem{Halperin_csitool}
D.~Halperin, W.~Hu, A.~Sheth, and D.~Wetherall, ``Tool release: Gathering
  802.11n traces with channel state information,'' \emph{ACM SIGCOMM CCR},
  vol.~41, no.~1, p.~53, Jan. 2011.

\bibitem{liu2015tracking}
J.~Liu, Y.~Wang, Y.~Chen, J.~Yang, X.~Chen, and J.~Cheng, ``Tracking vital
  signs during sleep leveraging off-the-shelf wifi,'' in \emph{Proceedings of
  the 16th ACM international symposium on mobile ad hoc networking and
  computing}, 2015, pp. 267--276.

\bibitem{7980063}
X.~Wang, C.~Yang, and S.~Mao, ``Phasebeat: Exploiting csi phase data for vital
  sign monitoring with commodity wifi devices,'' in \emph{2017 IEEE 37th
  International Conference on Distributed Computing Systems (ICDCS)}, 2017, pp.
  1230--1239.

\bibitem{hyvarinen2017nonlinear}
A.~Hyvarinen and H.~Morioka, ``Nonlinear ica of temporally dependent stationary
  sources,'' in \emph{Artificial Intelligence and Statistics}.\hskip 1em plus
  0.5em minus 0.4em\relax PMLR, 2017, pp. 460--469.

\bibitem{locatello2019challenging}
F.~Locatello, S.~Bauer, M.~Lucic, G.~Raetsch, S.~Gelly, B.~Sch{\"o}lkopf, and
  O.~Bachem, ``Challenging common assumptions in the unsupervised learning of
  disentangled representations,'' in \emph{international conference on machine
  learning}.\hskip 1em plus 0.5em minus 0.4em\relax PMLR, 2019, pp. 4114--4124.

\bibitem{mita2021identifiable}
G.~Mita, M.~Filippone, and P.~Michiardi, ``An identifiable double vae for
  disentangled representations,'' in \emph{International Conference on Machine
  Learning}.\hskip 1em plus 0.5em minus 0.4em\relax PMLR, 2021, pp. 7769--7779.

\end{thebibliography}
\end{document}